\documentclass[10pt,journal,compsoc]{IEEEtran}

\ifCLASSOPTIONcompsoc
  \usepackage[nocompress]{cite}
\else
  \usepackage{cite}
\fi

\usepackage{xcolor}

\usepackage{multirow}
\usepackage{subcaption}
\usepackage{caption}
\usepackage{graphicx}
\usepackage{booktabs}
\usepackage{xcolor}
\usepackage{mathtools}
\usepackage{amsfonts}
\usepackage{hyperref}

\usepackage{color, colortbl}
\usepackage{tabularx}
\usepackage{numprint}

\usepackage[first=0,last=9]{lcg}
\definecolor{Gray}{gray}{0.9}
\definecolor{White}{gray}{1}

\newcommand\sewaAU{SEWA AU DATASET}

\usepackage{pifont}
\usepackage{multirow}
\usepackage{array}
\usepackage{scrextend}
\usepackage{adjustbox}

\newcolumntype{L}[1]{>{\raggedright\let\newline\\\arraybackslash\hspace{0pt}}m{#1}}
\newcolumntype{C}[1]{>{\centering\let\newline\\\arraybackslash\hspace{0pt}}m{#1}}
\newcolumntype{R}[1]{>{\raggedleft\let\newline\\\arraybackslash\hspace{0pt}}m{#1}}
\newcolumntype{H}{>{\setbox0=\hbox\bgroup}c<{\egroup}@{}}

\newcommand{\eg}{e.\,g., }
\newcommand{\ie}{i.\,e., }

\begin{document}

\title{SEWA DB: A Rich Database for Audio-Visual Emotion and Sentiment
Research in the Wild
}
\author{Jean~Kossaifi,
        Robert~Walecki,
        Yannis Panagakis,
        Jie Shen,
        Maximilian Schmitt,
        Fabien Ringeval,\\
        Jing Han,
        Vedhas Pandit,
        Antoine Toisoul,
        Bj\"orn Schuller,~\IEEEmembership{Fellow,~IEEE},
        Kam Star,
		Elnar Hajiyev,
        and~Maja~Pantic,~\IEEEmembership{Fellow,~IEEE}

}

\IEEEtitleabstractindextext{
\begin{abstract}
    Natural human-computer interaction and audio-visual human behaviour sensing systems, which would achieve robust performance in-the-wild are more needed than ever as digital devices are increasingly becoming an indispensable part of our life. Accurately annotated real-world data are the crux in devising such systems. However, existing databases usually consider controlled settings, low demographic variability, and a single task. In this paper, we introduce the SEWA database of more than 2000 minutes of audio-visual data of 398 people coming from six cultures, 50\% female, and uniformly spanning the age range of 18 to 65 years old. Subjects were recorded in two different contexts: while watching adverts and while discussing adverts in a video chat. The database includes rich annotations of the recordings in terms of facial landmarks, facial action units (FAU), various vocalisations, mirroring, and continuously valued valence, arousal, liking, agreement, and prototypic examples of (dis)liking. This database aims to be an extremely valuable resource for researchers in affective computing and automatic human sensing and is expected to push forward the research in human behaviour analysis, including cultural studies. Along with the database, we provide extensive baseline experiments for automatic FAU detection and automatic valence, arousal and (dis)liking intensity estimation.
 \end{abstract}

\begin{IEEEkeywords}
SEWA, Affect Analysis In-the-wild, emotion recognition, regression.
\end{IEEEkeywords}}

\maketitle

\IEEEdisplaynontitleabstractindextext

\IEEEpeerreviewmaketitle

\IEEEraisesectionheading{\section{Introduction}\label{sec:introduction}}

\IEEEPARstart{A}{rtificial} Intelligence (AI) technologies are enabling the development of intelligent systems that are human-affect-aware and trustworthy, meaning that they can automatically detect and intelligently respond to users’ affective states \cite{brave2002emotion}.  Affecting computing requires methods that can robustly and accurately analyse human facial, vocal as well as verbal behaviour and interactions in-the-wild, that is, from data captured by omnipresent audio-visual sensors in digital devices in almost arbitrary recording conditions including semi-dark, dark and noisy rooms. 

Although notable progress has been made so far in machine analysis of human emotion and sentiment, there are still important challenges that need to be addressed in order to deploy and integrate affect-aware systems into everyday interfaces and real-world contexts. Concretely \cite{pantic2003toward,d2015review}: 
\begin{itemize}
    \item The vast majority of available datasets suitable for audio-visual emotion and sentiment research (cf. Section~\ref{sec:corpora} for a brief overview)  have been collected in laboratory or controlled conditions, with controlled noise level and reverberation, often limited verbal content, illumination and calibrated cameras. Such conditions are not present in real-world applications and tools trained on such data usually do not generalise well to behavioural recordings made in-the-wild.
    \item Many of the available datasets contain examples of induced behaviour as opposed to spontaneous behaviour (occurring in real-life settings of users’ natural environment like their home). As explained in various studies e.g., \cite{rinn1984neuropsychology,ambadar2009smiles}, spontaneous facial movements are smooth and ballistic, and are more typical of the subcortical system (not associated with cortex and displayed unconsciously). On the other hand, induced facial expressions may be planned and socially modified to a certain extent (i.e., associated with cortex and produced consciously), with entirely different dynamical characteristics than fully spontaneous facial expressions. Consequently, the dynamics of the behaviour
(timing, velocity, frequency, temporal inter-dependencies between gestures) crucially affect facial and vocal behaviour interpretation, and currently they are not taken into account.
	\item Observed behaviours may be influenced by those of an interlocutor and thus require analysis of both interactants, especially to measure such critically important patterns as mimicry, rapport or sentiment. However, existing approaches typically perform analysis of a single individual, and webcam-mediated face-to-face human computer interaction (FF-HCI) is not addressed as a problem of simultaneous analysis of both interacting parties.
    \item Available audio-visual databases are typically culture specific, e.g., the VAM faces database \cite{grimm2008vera} consists of 20 German speakers, the SEMAINE database consist of UK subjects \cite{mckeown2012semaine}, the RECOLA database consists of French speaking participants, the CONFER dataset contains only Greeks \cite{CONFER}. Hence, there is no database that would enable a large scale study on the effect of culture on expression recognition and communication of emotions and sentiment. 
    \item Most of the existing databases are only annotated in terms of certain behaviour and affect dimensions, for instance, the SEMAINE database contains continuous annotations of valence and arousal, the CONFER dataset is annotated only in terms of conflict intensity etc. Moreover, there is no database annotated in terms of multiple behavioural cues: facial action units (FAUs), affect dimensions and social signals.
\end{itemize}

In this paper, we aim to address the above mentioned challenges and limitations of existing datasets by introducing the SEWA database (SEWA DB), in Section~\ref{sec:sewadb}, which is an audio-visual, multilingual dataset of richly annotated facial, vocal, and verbal behaviour recordings made in-the-wild. The SEWA DB extends and contrasts considerably the available audio-visual datasets for affect research by providing the following key features:
\begin{itemize}
    \item The SEWA DB consists of audio-visual recordings of spontaneous behaviour of volunteers, captured in completely unconstrained, real-world, environments using standard web-cameras and microphones.
    \item It contains episodes of unconstrained interactions of subjects of different age, gender, and cultural backgrounds. In particular, $6$ groups of volunteers with around $66$ subjects per group (50\% females, uniformly divided over 5 age groups, 20+, 30+, 40+, 50+, 60+) from six different cultural backgrounds, namely British, German, Hungarian, Greek, Serbian, and Chinese were recoded. This makes the SEWA DB the first publicly available benchmark dataset 
for affect analysis in the wild across age and cultures.
  \item  Audio-visual recordings in the SEWA DB are richly annotated in terms of FAUs, facial landmarks, vocal and verbal cues as well as continuously valued emotion dimensions such as valence, arousal, liking and social signals including agreement and mimicry. This unique feature will allow for the first time to study different aspects of human affect simultaneously, investigate how observed behaviours are influenced in dyadic interactions, and exploit behaviour dynamics in affect modelling and analysis. Furthermore, the breadth of the annotations will allow to exploit inter-dependencies between age, gender, word and language usage, affect and behaviour, hence enabling robust and context-sensitive interpretation of speech and non-verbal behaviour.
\end{itemize}

For benchmarking and comparison purposes, we provide exhaustive baseline experimental results FAUs detection and valence, arousal and liking/disliking estimation are provided in Section~\ref{sec:exp}.

The SEWA database is available online at \url{http://db.sewaproject.eu/} and will not only be an extremely valuable resource for
researchers in affective computing and automatic human sensing but it
may also push forward the endeavour in human behaviour analysis,
especially when it comes to cross-cultural studies.  

\section{State-of-the-art in audio-visual emotion databases}
\label{Sec_SoA}

The standard approach in automatic emotion recognition relies on machine learning models trained on a collection of recordings, annotated in terms of different categories or dimensions of affect. As a consequence, the quality of the trained models, and especially their generalisation ability on new data acquired in different conditions, strongly depends on a myriad of factors that shape the construction of the emotional dataset itself. In this section, we discuss three of those main factors, namely elicitation methods, models of emotion representation, and data annotation techniques. In section \ref{sec:corpora}, we provide an overview of existing corpora.

\subsection{Elicitation methods}
One of the factors that has a significant impact on the models of emotion is the type of elicitation methods used to collect affective data. 
In order to record expressions of affect, one needs to consider a suitable context in which those expressions will be observed. 
Three main types of context have been used so far to collect such data: (i) \emph{posed behaviour} -- emotion is portrayed by a person upon request, \eg \cite{bilakhia2015mahnob,Banziger12-ITG}, (ii) \emph{induced behaviour} -- a controlled setting is designed to elicit a reaction to a given affective stimulus, \eg watching audio-visual clips or interacting with a manipulated system \cite{DISFA,Koelstra12-TGM}, and (iii) \emph{spontaneous behaviour} -- natural interactions between individuals, or between a human and a machine, are collected in a given context, \eg chatting with a sensitive artificial listener \cite{mckeown2012semaine}, or resolving a task in collaboration\cite{ringeval2013introducing}.

\subsubsection{Posed behaviour}
Interaction scenarios based on a posed behaviour present the advantage to know in advance the expressed emotion, since the portrayals are acted. 
Targeted emotions usually include the six ''basic emotions`` \cite{Ekman11-WIM}, and for which evidence for some universality over various cultures has been shown \cite{Ekman93-FEA}.
Acted scenarios further provide a fine-grained control of the material used to collect data, \eg phonetic complexity of the spoken utterances can be balanced for vocal analysis \cite{Banziger12-ITG}, as well as illumination or pose variations for facial analysis \cite{Gross10-MP}. 
In order to facilitate the portrayal of emotion, scripted scenarios can be exploited, with eventually the help of a professional director, who can interact with the actor, thus providing a more natural context \cite{Busso08-IIE, Banziger12-ITG}.

The automatic analysis of acted expressions of six basic emotions is now considered a solved problem with high accuracy performance reported in the literature \cite{Zeng09-ASO,gunes2013categorical,Ayadi11-SOS}.
Acted data can be of great interest when one wants to focus on specific details of emotional expressions. For instance, this can be very helpful for building rule-based prediction systems \cite{pantic2000expert,Bone14-RUA}, or when the targeted population presents major difficulties in handling complex display of emotion, such as in the autism spectrum conditions \cite{Golan15-TCM, Marchi15-AAS}.
On the other hand, acted data cannot be used for training when one wants to predict natural display of emotion. Spontaneous expressions are much more subtle in comparison with acted portrayals. As a consequence, they are also much more challenging to recognise \cite{Schuller11-RRE}.

\subsubsection{Induced behaviour}
In order to collect naturalistic expressions of emotion, one can induce affect by using either passive or active methods. Passive methods consist in (dis)playing a set of standardised stimuli to subjects whose reactions (vocal, facial, and physiological) are recorded. 
Stimuli can be either static, \eg the International Affective Picture Systems (IAPS) \cite{lang2007international}, or dynamic, \eg audio clips \cite{kim2008emotion}, video clips \cite{mahnob_laughter_db} or excerpts from movies \cite{Dhall12-CLR}. They can also be incorporated into a human-computer interaction system, in order to provoke affective reaction from the user. For instance, system malfunctions or unexpected events can be generated automatically, or by a Wizard-of-Oz, in order to induce emotion \cite{walter2013transsituational}. 
Induced behaviours are also of interest for emotionally driven marketing research, e.g., the efficiency of an audiovisual advertisement can be measured automatically through the affective reactions of the audience, instead of self-reported questionnaires.

\subsubsection{Spontaneous behaviour}
The most appealing approach for capturing a wide range of fully naturalistic displays of emotions consists in recording spontaneous human interactions.
Ecologically valid situations, \ie observing humans in their natural environments, would be ideal as it ensures unobtrusiveness and thus guarantees the observation of fully natural behaviours.
This step out of the laboratory has not been accomplished until now for the collection of affective data produced during human interactions. The SEWA database presented in this paper is the very first such collection of interactive human behaviour, annotated in terms of displayed affective dimensions, recorded in-the-wild. Until now, various characters presenting different personality traits (\eg joyful, depressed, introvert, etc), and simulating an artificial sensitive listener, have served as human interlocutors \cite{mckeown2012semaine}.

\subsection{Representation of emotion}
Emotion is a subjective feeling, and a complex internal phenomenon. Hence, using a simplistic classification-based model relying on few emotion categories provides only a very limited description of the phenomenon. Additionally, whether or not a certain expression stems from one emotion or the other (\eg sadness versus boredom) could be matter of subjective interpretation. 

In the pursuit of a finer model, several continuous-valued, multidimensional models have been proposed for more precise emotion description. Arguably, the most popular model employed by the affective computing research community is the two dimensional model describing the degree of activation (arousal) and pleasantness (valence) of displayed affect expressions as a point in the Cartesian plane. A well known problem with this approach is the dynamically varying time-delay between an expression and the annotation due to reaction lag of the annotators, which also varies among different annotators, over the sessions, and even during every session \cite{nicolaou_tpamiDPCCA}. Yet methods have been proposed for spatio-temporal alignment of annotations \cite{nicolaou_tpamiDPCCA,GCTW} to remedy this problem and come up with reliable ground truth to be used for training dimensional affect regressors.

\subsection{Data Annotation and generation of the Gold Standard}

Depending on the choice of the model for emotion representation, several research groups have developed their own annotation tools. Some of these tools have been now made available to researchers across the world, some even with open source licenses. Popular annotation tools in use today are ANVIL \cite{kipp2001anvil}, ATLAS \cite{meudt2012atlas}, Ikannotate \cite{bock2011ikannotate}, EmoWheel (Geneva emotion wheel) \cite{scherer2010developing}, FEELtrace \cite{cowie2000feeltrace}, Gtrace \footnote{Successor to FEELtrace, \url{https://sites.google.com/site/roddycowie/work-resources}}, ANNEMO \cite{Ringeval13-ITR}, and the frame by frame Valence/Arousal Online Annotation Tool \cite{AFEWVA_kossaifi}.

Several methods exist for creating a unified view of the perceived emotion from a set of annotations, generally referred as the `Gold Standard' to differentiate with `ground-truth', which is avoided for affective computing as there does not exist a truth on a subjective feeling such as emotion. 
The basic principle is to use consensus among the evaluators to come up with a common, best representative annotation by using different metrics such as the correlation coefficients, dynamical time warping (DTW) distance \cite{RCICA,GCTW}, average of the data post standardization or normalization, or assigning individual annotations certain weight percentages (\eg evaluator weighted estimator (EWE)).

 \subsection{The Existing Corpora}
\label{sec:corpora}

We focus here on databases containing dyadic interaction recordings annotated in terms of displayed affective reactions. For an overview of databases containing recordings of non-interactive subjects, the reader is referred to recent survey papers (e.g. \cite{gunes2013categorical,automatic_survey_2015}) and recent database papers (e.g. \cite{AFEWVA_kossaifi}). Most of the databases of dyadic interaction recordings annotated in terms of displayed affective reactions contain recordings made in controlled settings, concern a constrained dyadic task, have low demographic variability, and are made in primarily one language --that has mostly been English so far. 
Predominance of one language in the corpora limits usability of the database for cross-lingual, cross-cultural study of emotion recognition. 
Table~\ref{Tab_DBcompare} presents a summarized overview of the surveyed databases containing dyadic interactions.

\setlength{\tabcolsep}{0.4em}
\begin{table*}[!th]
\centering
	\caption{Summary of corpus available for estimation of  arousal and valence from audiovisual data, featuring unscripted interactive discourse. 
    Information that is not available in the citation is indicated as 'NI', short for `No Information'.
    }
\label{Tab_DBcompare}
\setlength\tabcolsep{0pt}
\setlength\extrarowheight{2pt}
\begin{tabular*}{1\textwidth}{@{\extracolsep{\fill}} L{2cm}rrrL{1.5cm}HL{1.5cm}L{2.5cm}crR{1.8cm}Hr}
\toprule	
\multirow{2}{*}{\textbf{Dataset}} 
	& \multicolumn{4}{c}{\textbf{\#Subjects}}
        & \multirow{2}{*}{\textbf{\begin{tabular}[c]{@{}c@{}} Balanced \\Age Diversity\end{tabular}}}
    & \multirow{2}{*}{\textbf{\begin{tabular}[c]{@{}c@{}}\# audio-\\-visuals 
    \end{tabular}}}
	& \multirow{2}{*}{\textbf{Annotation}}
	& \multirow{2}{*}{\textbf{Duration}} 
	& \multirow{2}{*}{\textbf{Language(s)}}
	& \multirow{2}{*}{\textbf{Elicitation}}
    & \multirow{2}{*}{\textbf{Illumination}} 
    & \multirow{2}{*}{\textbf{Year}} 
    \\
    \cmidrule{2-5}
    & \textbf{Total}
	& \multicolumn{1}{c}{\textbf{M}}
	& \multicolumn{1}{c}{\textbf{F}}
	& \multirow{1}{*}{\textbf{\begin{tabular}[c]{@{}c@{}} Age range\end{tabular}}}
	&
	&	\\\hline
\begin{tabular}[c]{@{}l@{}}GENEVA \cite{scherer1997lost}
\end{tabular}
	& {112} 	& 55	& 45 &\begin{tabular}[c]{@{}c@{}}20-60+\end{tabular} & NI
    & 112
    & NI
	& NI
	& \begin{tabular}[c]{@{}r@{}}French\\English\\German/\\\hspace{0.1cm}northern Europe\\Asia\\Other \end{tabular}
	& \begin{tabular}[c]{@{}c@{}}conversation 
    \end{tabular}
	& controlled  
    & 1997
    \vspace{0.1cm}
    \\ 
\begin{tabular}[c]{@{}l@{}}SMARTKOM \cite{schiel2002smartkom}  \end{tabular}

	& {224} 	&	NI		& NI  &16-45+ & No
    & 466
    & NI
	& 17 hours 
	& \begin{tabular}[c]{@{}l@{}}German\end{tabular}
	& HMI 
	& controlled   
    & 2002
    \vspace{0.1cm}
    \\ 
 VAM-faces \cite{grimm2008vera}
	& {20} 	& NI & NI  &\begin{tabular}[c]{@{}c@{}}16-69\\ 70\%\textless35\end{tabular} & No
    & 1421
    & Linkert-like scale (5 points from -1 to 1), 7--8 raters
	& 12 hours
	& \begin{tabular}[c]{@{}l@{}}German\end{tabular}
	& Talk-show 
	& controlled   
    & 2008
    \\
SEMAINE \cite{mckeown2012semaine}
	& {150} &	57	& 93 & NI & No 
    & 959
    & continuous, Feel-trace, 7 raters
	& 80 hours
	& \begin{tabular}[c]{@{}l@{}}English\end{tabular}
	& HMI
	& controlled  
    & 2012
    \vspace{0.1cm}
    \\ 
Belfast naturalistic \footref{BNDf}
	& 125 	& 31		& 94 & NI & No
    & 298
    & Continuous, Feeltrace, 6--258 raters
	& 86 minutes
	& \begin{tabular}[c]{@{}l@{}}English\end{tabular}
	& Talk-show
	& controlled  \vspace{0.1cm}
    & 2010 \\ 
  AVEC'13 \cite{Valstar13-A2T}
	& {292} &	NI	& NI & 18-63 & No 
    & 340 
    & Continuous, Feeltrace, 1 rater
	& 240 hours
	& \begin{tabular}[c]{@{}l@{}}German\end{tabular}
	& HMI
	& controlled  
    & 2013
    \vspace{0.1cm}
    \\ 
Belfast induced 1 \cite{sneddon2012belfast}
	& {114} 	& 70		& 44 & \multirow{2}{*}{NI} & \multirow{2}{*}{No} 
    & 570
    & Continuous, Feeltrace, 6--258 raters
	& 237 minutes
	& \multirow{2}{*}{\begin{tabular}[c]{@{}l@{}}English\end{tabular}}
	& TV/interviews
	& \multirow{2}{*}{controlled} 
    & \multirow{2}{*}{2012}
    \\ 
Belfast induced 2 \cite{sneddon2012belfast}
    & {82} 	& 37		& 45 & NI & 
    & 650 
    & Valence only, Continuous, Feeltrace, 1 rater
	& 458 minutes
	& 
	& laboratory based tests
	& 
    \vspace{0.1cm}
    \\ 
CCDb \cite{aubrey2013cardiff}
	& 16 	& 12		& 4 & 25-56 & No
    & 30
    & NI
	& 300 minutes
	& \begin{tabular}[c]{@{}l@{}}English\end{tabular}
	& conversations 
	& controlled   
    & 2013
    \vspace{0.1cm}
    \\

RECOLA \cite{ringeval2013introducing}
	& {46} &	19	& 27 & NI & no 
    & 46
    & Continuous, Feeltrace, 7 raters 
	& 230 minutes
	& \begin{tabular}[c]{@{}l@{}}French\end{tabular}
	& online conversation
	& controlled   
    & 2013
    \vspace{0.1cm}
    \\ 
AVEC'14 \cite{Valstar14-A2T} 
	& {84} &	NI	& NI  & 18-63 & No 
    & 300 
    & Continuous, Feeltrace, 3+ raters
	& 240 hours
	& \begin{tabular}[c]{@{}l@{}}German\end{tabular}
	& HMI
	& controlled   
    & 2014
    \vspace{0.1cm}
    \\
MAHNOB Mimicry \cite{bilakhia2015mahnob}
	& {60} & 31 & 29  & NI & No 
    & 54
    & Continuous, Feeltrace, $\simeq$ 5 raters
	& 11 hours
	& \begin{tabular}[c]{@{}l@{}}English\end{tabular}
	& dyadic conversations
	& controlled 
    & 2015
    \vspace{0.1cm}
    \\
4D CCDb \cite{vandeventer20154d}    
	& {4} &	2	& 2 & 20-50 & No
    & 34 
    & NI
	& 17 minutes
    & \begin{tabular}[c]{@{}l@{}}English\end{tabular}
	& conversation
	& controlled  
    & 2015
    \vspace{0.1cm}
    \\

\midrule   
\textbf{SEWA (this work)}
	& {398} &	201	& 197 &  18-60+ & No
    & 1990
    & Continuous, Feeltrace, 5 raters
	& 44 hours
	& \begin{tabular}[c]{@{}r@{}}Chinese\\English\\German\\Greek\\Hungarian\\Serbian\end{tabular}
	& Watching videos -- Dyadic conversations
	& uncontrolled  
    & 2017
    \vspace{0.1cm}
    \\ 
\bottomrule 
\end{tabular*}
\end{table*}
 
\subsubsection{Elicitation using Conversational Context}
The \emph{Geneva Airport Lost Luggage Study} database \cite{scherer1997lost} is amongst the very few  databases featuring cultural diversity in terms of its subjects. 
112 passengers --that were required to claim their lost baggage at the airline's baggage retrieval office, were recorded surreptitiously during their interaction with the airline agents processing their claims. Because of this unobtrusive recording paradigm, the dataset features truly natural emotional responses, and can be claimed to be free from Labov's paradox \cite{labov1972sociolinguistic}. The gender split of the data is adequately balanced with 59.8\,\% male, and 40.2\,\% female subjects. The linguistic/cultural distribution however is quite unbalanced. In addition, individual languages of the participants is not reported, only 'language groups' are given \cite{scherer1997lost}. 
No continuous annotations are available, only subjects overall feeling for the whole episode are provided. Specifically, the subjects self-reported their emotional states before and after the interaction in a 7-point scale for 5 emotion categories; namely 
`angry/irritated', `resigned/sad', `indifferent', `worried/stressed', and `in good humour'. 

The \emph{Cardiff Conversation Database (CCDb)} \cite{aubrey2013cardiff} and \emph{4D Cardiff Conversation Database (4D CCDb)} \cite{vandeventer20154d} databases follow another interesting recording paradigm where the subjects 
freely discuss topics of their own interest and lead the conversations themselves.
They are not given any specific task, nor a topic for conversation, nor any specific audiovisual stimuli to elicit emotions.  
Both the databases contain conversations only in English, each containing 30 and 6 conversations respectively. The gender split is quite skewed with 12 male and only 4 female subjects. The dataset is annotated in terms of Frontchannel (main speaker periods), Backchannel (qualified utterances and expressions), agreement/disagreement episodes, smiles, laughter, negative and positive surprises, thinking phases, confusion, head motions, made using ELAN \cite{wittenburg2006elan} framework.

The \emph{MAHNOB Mimicry} \cite{bilakhia2015mahnob} dataset features dyadic conversations where subjects engage in socio-political discussions, or negotiate a tenancy agreement. The subjects span range of nationalities including Spanish, French, Greek, English, Dutch, Portuguese,and Romanian. All conversations were recorded in English. Subjects are 18 to 34 years old, with 4.8 years of standard deviation. Continuous annotations were obtained using FeelTrace by approximately 5 raters.

The \emph{Conflict Escalation Resolution (CONFER)} \cite{CONFER} dataset is constituted of 120 video clips of interactions between 54 subjects from Greek televised political debates. The data was annotated by 10 experts in terms of continuous conflict intensity.

\subsubsection{Elicitation using Human-Machine Interfaces}
The \emph{Sustained Emotionally coloured Machine-human Interaction using Nonverbal Expression Dataset (SEMAINE)} \cite{mckeown2012semaine}  presents richly annotated recordings  (7 basic emotion states, 6 types of epistemic states, transcripts, laughs, head movement and FACS) of interactions in laboratory conditions between a human and a machine-like agent in three different Sensitive Active Listener (SAL) scenarios. It features 150 participants, most of which come from Caucasian background and 38\,\% are male. The language of communication is predominantly English.

The \emph{SMARTKOM} dataset \cite{schiel2002smartkom} features subjects interacting in laboratory conditions, in German with a pretense/WOZ multimodal dialogue system that supposedly allows the  user to interact almost naturally with a computer. 
The recording sessions were first split into subject-state episodes by the labellers marking start and end of each perceived episode. The segments were then labelled with the following 7 categories: `joy/gratification', `anger/irritation',  `helplessness', `pondering/reflecting',  `surprise', `neutral' and `unidentifiable episodes'. Gender split is 20 male and 25 female speakers.

\subsubsection{Elicitation through Tasks}
The {RECOLA} \cite{ringeval2013introducing} dataset contains multimodal recordings of French students performing a collaborative task. 
The participants discuss and rank 15 items in the order of their significance for their survival in a remote and hostile region in cold winter. The subjects are from different parts of Switzerland, and thus have different cultural backgrounds (33 French, 8 Italian, 4 German, 1 Portuguese). The database however features French language alone, and mean age of the subjects is 22 years with only 3 years of standard deviation. Continuous levels of valence and arousal were annotated by 7 raters using FeelTrace.

To collect \emph{Belfast Induced Natural Emotion Database} \cite{sneddon2012belfast}, English speaking participants were asked to perform select set of tasks specifically designed to induce mild to moderately strong emotionally coloured responses (e.g. reaching into a box that sets off a very loud alarm). Mean age of subjects is 24 years with 6 years of deviation. Continuous values of valence and arousal were obtained for each clip by 6 to 258 raters using FeelTrace.

The corpus for Audio-Visual Emotion recognition Challenges in 2013 and 2014, namely \emph{AVEC'13} \cite{Valstar13-A2T} and \emph{AVEC'14} \cite{Valstar14-A2T}, used a subset of audio-visual depressive language corpus (AViD-Corpus) which consists of recordings of subjects performing human-computer interaction tasks, labelled by 23 annotators continuous for arousal and valence estimates. The mean age of subjects is 31 years, with 6 years standard deviation.

\subsubsection{Corpus collected by segmenting existing recordings}
\emph{Belfast  Naturalistic Database} \footnote{\url{http://sspnet.eu/2010/02/belfast-naturalistic/}\label{BNDf}} contains 10 to 60 seconds--long audiovisuals taken from English television chat shows, current affairs programmes and interviews. It features 125 subjects, of which 31 are male, and 94 are females. Out of 298 clips, 100 videos totalling 86 minutes in duration have been labelled with continuous-valued emotion labels for activation and evaluation dimensions, with additionally 16 basic classifying emotion labels. 

Similarly, \emph{Vera am Mittag (VAM) database} \cite{grimm2008vera} contains 12 hours of recordings of the German TV talk-show “Vera am Mittag” (Vera at noon) with continuous-valued emotion labels for arousal, valence, and dominance. It contains 20 participants with age ranging from 16 to 69.

The \emph{AFEW-VA database} \cite{AFEWVA_kossaifi} is a visual dataset containing \(600\) challenging video clips extracted from feature films and annotated per-frame in term of levels of valence and arousal, as well as \(68\) facial landmarks. It contains 240 subjects, 50\% female, with age ranging from 8 to 76.
 \section{SEWA database}
\label{sec:sewadb}
	The main aim of the SEWA DB is to provide enough suitable data of labelled examples to facilitate the development of robust tools for automatic machine understanding of human behaviour. 
	In this section, we discuss in detail 
	the steps we took to 
	collect both the audiovisual recordings and the corresponding diverse set of annotations. We then present the statistics pertaining to the collected recordings, and the link to the web-resource with a comprehensive search filter featuring all of the data we gathered.

\subsection{Data collection}
     To create the SEWA dataset, a data collection experiment has been conducted. In this experiment, participants were divided into pairs based on their cultural background, age and gender. During initial sign-up, participants were asked to complete a questionnaire of demographic measures including gender, age, cultural background, education, personality traits, and familiarity with the other person in the pair. To promote natural interactions, participants within each pair were required to know each other personally in advance of the experiment. Each pair of the participants then took part in two parts of the experiment, resulting in two sets of recordings.

    The SEWA data collection experiment was conducted using a website specifically built for this task (shown in Figure \ref{fig:annotation_website}). The website (\url{http://videochat.sewaproject.eu}) utilises WebRTC/OpenTok to facilitate the playing of adverts, video-chat, and synchronized audio/video recording using the microphone and webcam on the participants’ own computer. This setup allowed the participants to be recorded in truly unconstrained “in-the-wild” environments with various lighting conditions, poses, background noise levels, and sensor qualities.

\textbf{Experimental Setup Part 1: } Each participant was asked to watch 4 adverts, each of being around 60 seconds long. These adverts had been chosen to elicit mental states including amusement, empathy, liking and boredom. For consistent understanding of the advertisement content across cultures, the advertisements chosen have no dialogues, but are driven primarily by the visuals and the accompanying music. The four advertisements \footnote{These videos, including all subtitled versions prepared for subjects of different cultural backgrounds, are included in our dataset for referencing purposes. } in order are:

\begin{enumerate}
    \item 
    A violent advertisement of the \emph{National Domestic Violence Hotline} 
    --eliciting disgust, distress, and yet the liking for the effectiveness of the advertisement,
    \item 
    A self-deprecating, witty advertisement of the \emph{Smart Fortwo} car 
    --eliciting pleasure and liking for the advertisement, 
    \item 
    A bizarre, abstract advertisement of the \emph{Jean Paul Gaultier Le Male Terrible} perfume with a highly blurred product emphasis, with illegible product name in the visuals 
    --eliciting confusion and a strong disliking for the advertisement,
    \item 
    An advertisement of a touch-activated \emph{Grohe} faucet presenting use cases for the newly introduced sensor-based activation feature, 
    --eliciting interest and liking for the product, and boredom for the advertisement overall.
\end{enumerate}

After watching the advert, the participant was also asked to fill-in a questionnaire to self-report his/her emotional state and sentiment toward the advert.\\

\textbf{Experimental Setup Part 2: } After watching the 4th advert, the two participants were asked to discuss the advert they had just watched by using the video-chat function provided by the SEWA data collection website. On average, the recorded conversation was 3 minutes long. The discussion was intended to elicit further reactions and opinions about the advert and the advertised product, such as whether the advertised is to be purchased, whether it is to be recommended to others, what are the best parts of the advert, whether the advert is appropriate, how it can be enhanced, etc.. After the discussion, each participant was asked to fill-in a questionnaire to self-report his/her emotional state and sentiment toward the discussion.\\

\begin{figure}[t!]
    \centering
    \includegraphics[width=1.\linewidth]{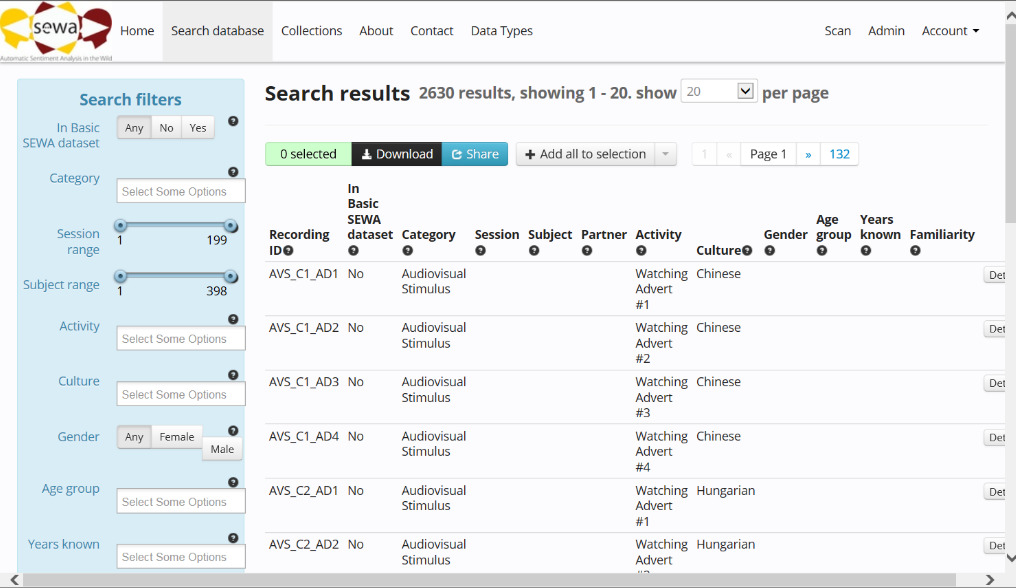}
    \caption{Front page of the online SEWA database and the search filters.}
    \label{fig:annotation_website}
\end{figure}

\subsection{The Data Statistics and subject demographics}
 \setlength{\tabcolsep}{0.1em}
\begin{table}[]
\centering
\caption{SEWA Demographics}
\label{Tab_SEWAsubjects}
\begin{tabular*}{1\columnwidth}{@{\extracolsep{\fill}} lcrrrrrr}
\toprule
\multicolumn{2}{l}{\multirow{2}{*}{}} & \multicolumn{6}{c}{Cultures}                           \\ \cmidrule{2-8}
\multicolumn{2}{l}{}                 & Chinese & English & German & Greek & Hungarian & Serbian \\\midrule
\multirow{2}{*}{Gender}    & Male    & 36      & 33      & 39     & 34    & 26        & 33      \\
                           & Female  & 34      & 33      & 25     & 22    & 44        & 39      \\\midrule
\multirow{3}{*}{Interactions}       
                           & F--F    & 22      & 20      & 16     & 8     & 30        & 16      \\
                           & M--F    & 24      & 26      & 18     & 26    & 28        & 46      \\
                           & M--M    & 24      & 20      & 30     & 22    & 12        & 10      \\\midrule
\multirow{5}{*}{Age}       & 18--29  & 44      & 34      & 41     & 18    & 44        & 22      \\
                           & 30--39  & 16      & 12      & 13     & 29    & 9         & 15      \\
                           & 40--49  &  4      & 6       & 1      & 1     & 5         & 8       \\
                           & 50--59  &  6      & 8       & 5      & 8     & 5         & 14      \\
                           & 60+     &  0      & 6       & 4      & 0     & 7         & 13      \\\midrule
\multicolumn{2}{c}{\textbf{Total}}
& 70 & 66 &     64 & 56 & 70 & 72                  \\
\bottomrule
\end{tabular*}
\end{table}     During the SEWA experiment, 198 recording sessions have been successful, with a total of 398 subjects being recorded. The subjects come from 
6 different cultural backgrounds: British, German, Hungarian, Serbian, Greek, and Chinese.
201 of the participants are male, 197 are female, resulting in a gender ratio (male / female) of 1.020. Furthermore, the participants are categorized into 5 age groups: 18~29, 30~39, 40~49, 50~59 and 60+, with the 18~29 group being most numerous. The detailed participant demographics are shown in Table \ref{Tab_SEWAsubjects}.

A total of 1990 audio-visual recording clips (5 clips per subject: 4 recorded during the advert-watching part and 1 recorded during the video-chat part) were collected during the experiment, comprising of 1600 minutes of audio-visual data of people's reaction to adverts and 1057 minutes of video-chat recordings. Due to the wide spread of the participants’ computer’s hardware capacity, the quality of the video and audio recordings is not constant. Specifically, the spatial resolution of the video recordings ranges from 320x240 to 640x360 pixels and the frame rate is between 20 and 30 fps. The audio recording’s sample rate is either 44.1 or 48 kHz. 

\subsection{Data annotation}

    The SEWA database contains annotations for facial landmarks, (pre-computed) acoustic low-level descriptors (LLDs) \cite{Schuller13-TI2} \cite{Eyben16-TGM}, hand gestures, head gestures, facial action units, verbal and vocal cues, continuously-valued valence, arousal and liking / disliking (toward the advertisement), template behaviours, episodes of agreement / disagreement, and mimicry episodes. 

    Due to the large amount of raw data acquired from the experiment, the annotation process has been conducted iteratively, starting with sufficient amount of examples to be annotated in a semi-automated manner and used to train various feature extraction algorithms developed in SEWA. Specifically, 538 short (10-30s) video-chat recording segments were manually selected to form the fully-annotated \textbf{basic SEWA dataset}. These segments were selected based on the subjects’ to the subjects’ emotional state of low / high valance, low / high arousal, and liking / disliking. All 6 cultures were evenly represented in the basic SEWA dataset, with approximately 90 segments selected from each culture based on the consensus of at least 3 annotators from the same culture.

\subsubsection{Facial landmarks}

Facial landmarks were annotated for all segments included in the basic SEWA dataset using a 49-point mark-up, as described in \cite{shen2015first}. Manual annotation of facial landmarks is highly labour intensive. Based on previous experience \cite{shen2015first}, we know that trained annotators can only achieve a sustained annotation speed of 30 frames per hour. Since the basic SEWA dataset contains a total of 369974 video frames, it would be impractical to annotate all of them manually (which would require more than 12000 hours of work). Therefore, the annotation was performed semi-automatically.

\begin{figure}[t!]
    \centering
    \includegraphics[width=1.\linewidth]{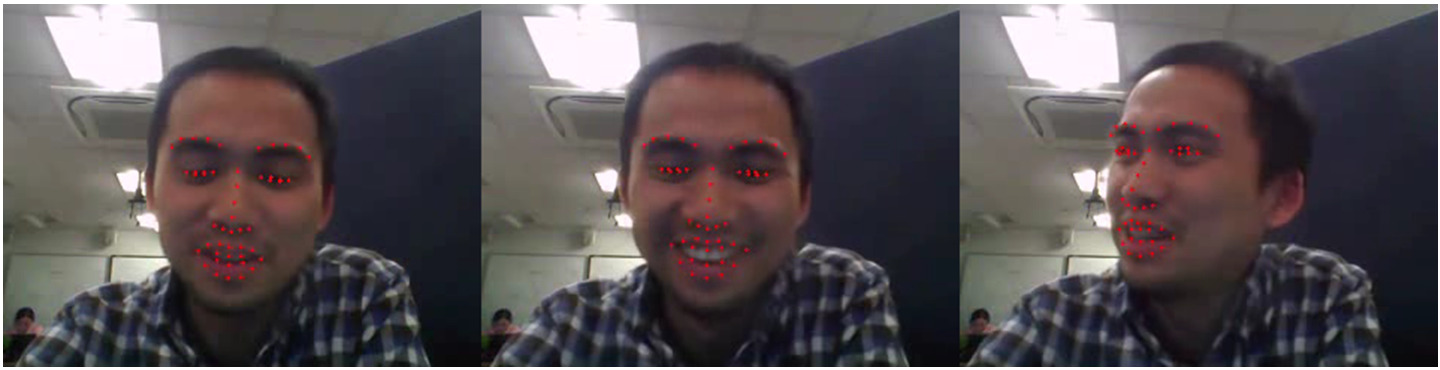}
        \caption{Example of facial landmark annotation. The 49 facial landmarks were annotated for all segments included in the basic SEWA dataset.}
    \label{fig:landmarks}
\end{figure}

During the annotation process, we first applied the Chehra facial landmark tracker \cite{asthana2014incremental}\cite{asthana2015pixels} on all video segments. Using a discriminative model trained by a cascade of regressors, the tracker can construct personalised model by incremental updating of the generic model. More than 95.1\% of the tracking results (in 351875 frames) produced by Chehra are accurate and require no further correction. For the remaining 18099 frames, manual annotation was performed in a similar way as in preparation of the 300VW dataset \cite{shen2015first}\cite{chrysos2015offline}. Specifically, we manually annotated 1 in every 8 frames and used the results to train a set of person-specific trackers. These person-specific trackers were applied to the rest of the frames to obtain the annotations. Finally, a visual inspection was performed on the annotations and those deemed unsatisfactory were further corrected. An example of the facial landmark annotation obtained from this process is show in Figure \ref{fig:landmarks}.

\begin{figure}[t!]
    \centering
    \includegraphics[width=1.\linewidth]{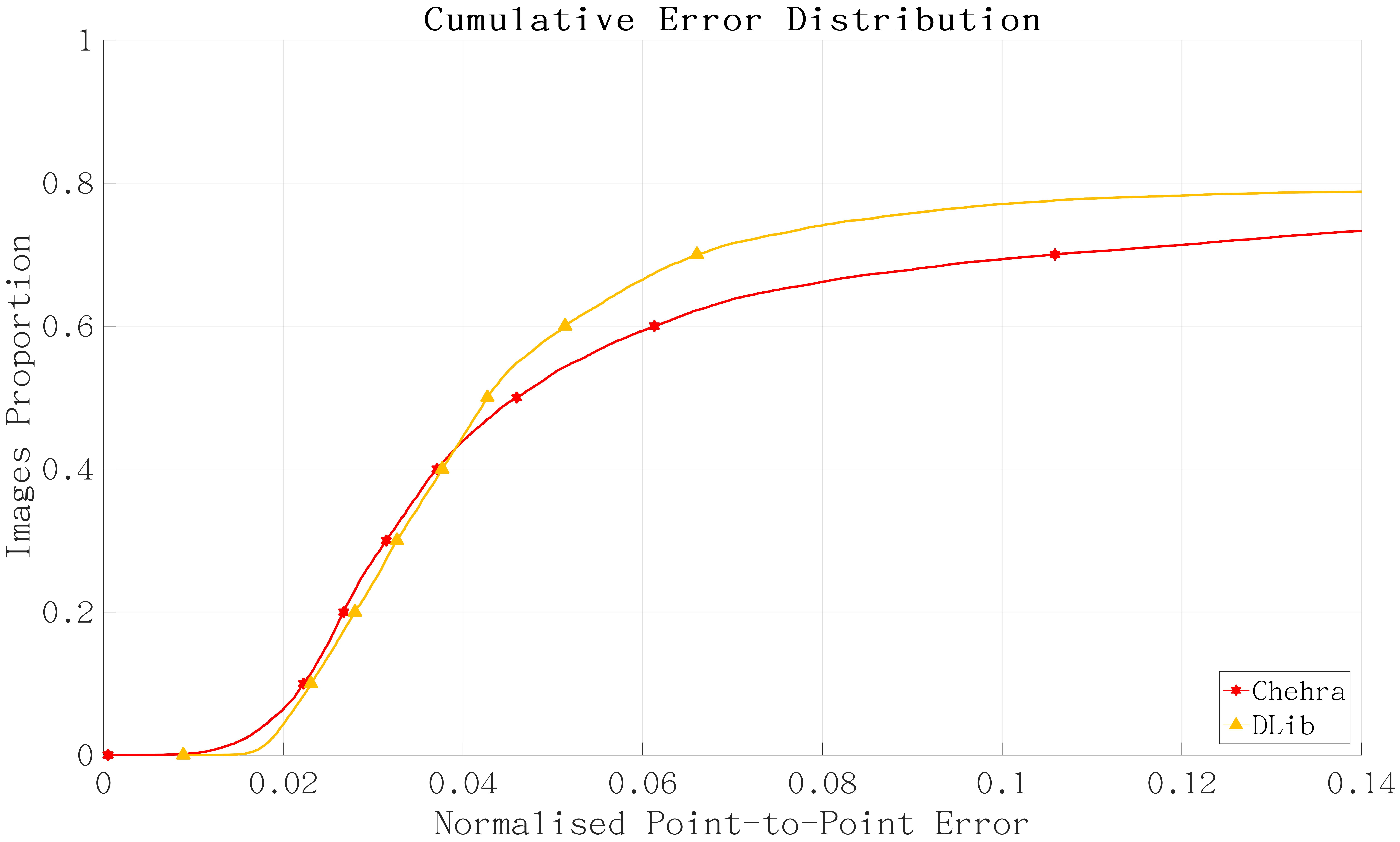}
        \caption{CED curves of the Chehra tracker \cite{asthana2014incremental} and the Dlib tracker \cite{kazemi2014one} on the manually corrected frames.}
    \label{fig:landmarks_baseline}
\end{figure}

\subsubsection{Hand Gesture}
\begin{figure}[t!]
    \centering
    \includegraphics[width=1.\linewidth]{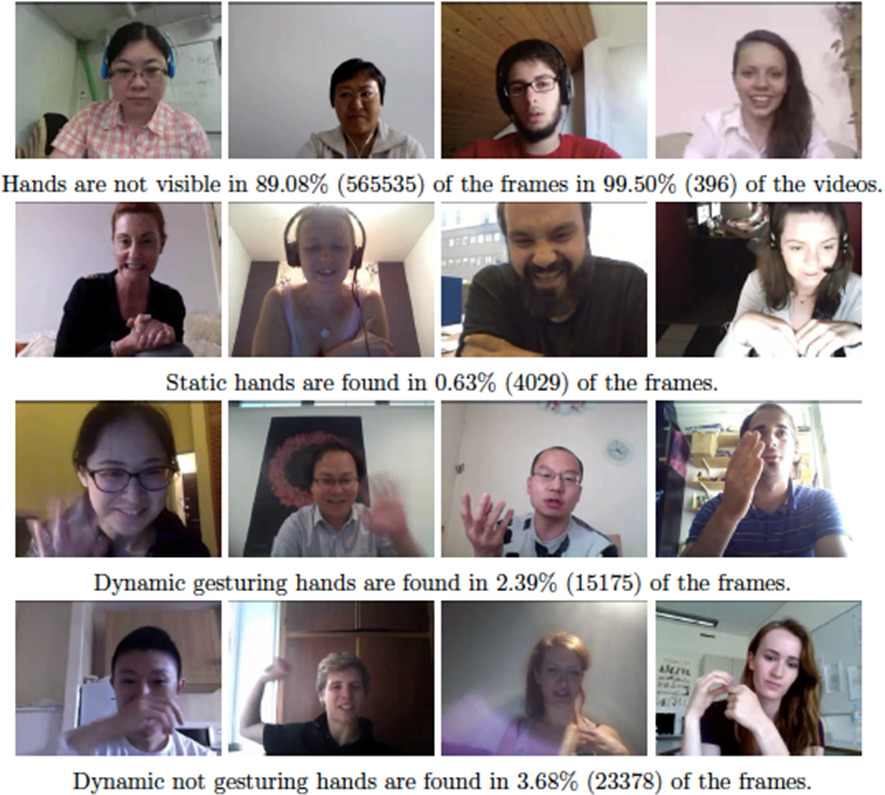}
        \caption{Examples of hand gesture annotation.}
    \label{fig:hand_gesture}
\end{figure}

Hand gestures were annotated for all video-chat recordings in 5 frame steps. Five types of hand gestures were labelled: hand not visible (89.08\%), hand touching head (3.32\%), hand in static position (0.63\%), display of hand gestures (2.39\%), and other hand movements (3.68\%). Some examples of the labelled frames are shown in Figure \ref{fig:hand_gesture}.

\subsubsection{Head Gesture}
\begin{figure}
    \centering
    \includegraphics[width=1.\linewidth]{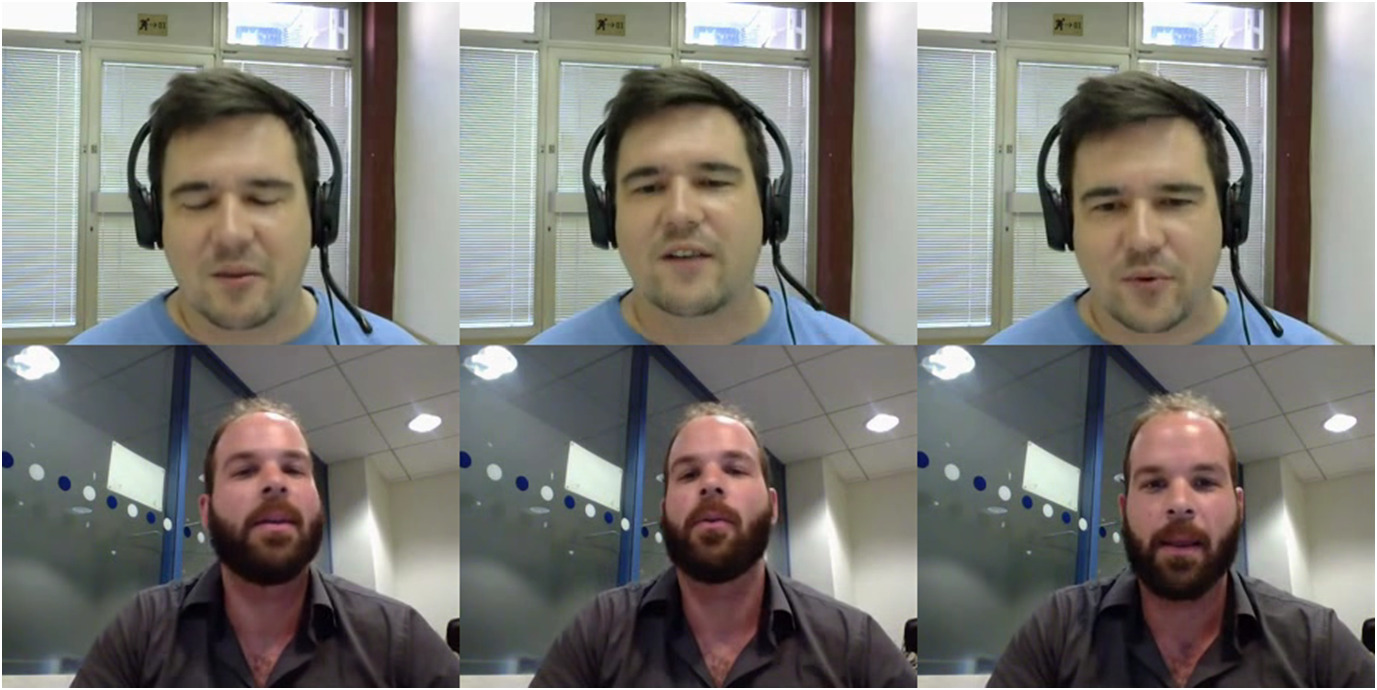}
    \caption{Examples of head nod (top row) and head shake (bottom row) sequences.}
    \label{fig:head_gesture}
\end{figure}

Head gestures were annotated in terms of nod and shake for all segments in the basic SEWA dataset. The annotation was performed manually on a frame-by-frame basis. To be able to provide good training examples for the head nod/shake detector, we emphasised specifically on high precision during the annotation process. Specifically, only un-ambiguous displays of head nod / shake were labelled. In the end, a total of 282 head nod sequences and 122 head shake sequences were identified. Examples of the labelled head nod / shake sequences are shown in Figure \ref{fig:head_gesture}

\subsubsection{Transcript}
We provide the audio transcript of all video-chat recordings. In addition to the verbal content, the transcript also contains labels of certain non-verbal cues, such as sighing, coughing, laughter, etc. Utterances were transcribed lexically, including markers for non-linguistic vocalizations including laughs, "back-channel" expressions of consent and hesitation, rapid audible nasal exhalation (like in smirk and snigger), and audible oral exhalations. To minimise the efforts for transcription, semi-automatic methods such as active learning were in an iterative fashion, starting with existing modules for automatic speech recognition and spotting of non- linguistic vocalisations \cite{wollmer2011robust}\cite{weninger2011munich}. Since lexical transcription does not require special training, crowd-sourcing was used.

\subsubsection{Facial Action Units annotation}

\begin{figure}
    \centering
    \includegraphics[width=1.\linewidth]{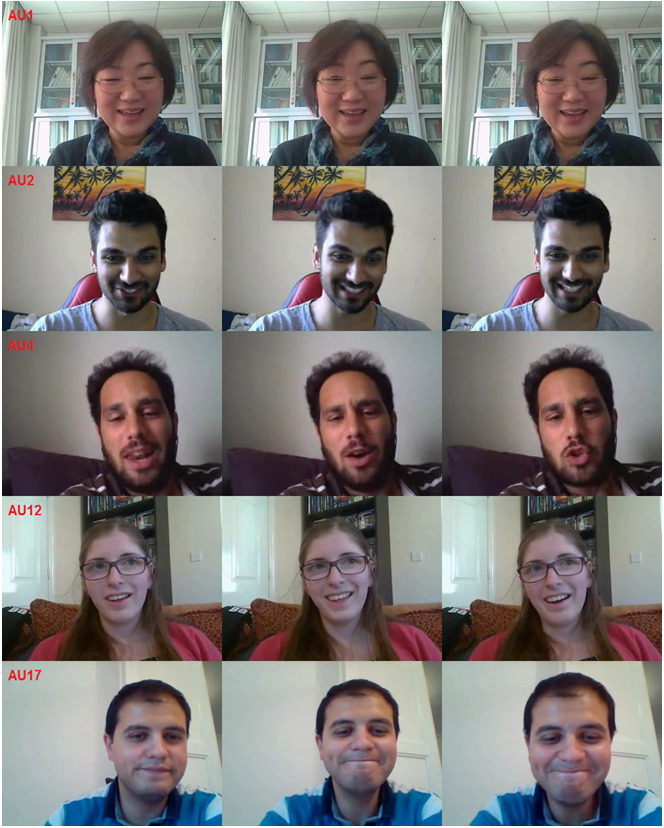}
    \caption{Examples of AU1 (inner brow raiser), AU2 (outer brow raiser), AU4 (Brow lowerer), AU12 (Lip corner puller) and AU17 (Chin raiser)}
    \label{fig:AUs}
\end{figure}

    Manual annotation of Action Units (AUs) has to be performed by trained experts which is expansive and time consuming. Especially due to the size of the SEWA database, such manual annotation is prohibitive. Therefore, we focus on accurate semi-automatic annotation of five AUs (1,2,4,12 and 17, depicted in Fig. \ref{fig:AUs}). We selected these AUs as they are occurring most in naturalistic settings, and are important for high-level reasoning about sentiment.

We leverage the state-of-the-art method of \cite{walecki2015variable} for detection of AUs. Specifically, we employ publicly available datasets (DISFA \cite{mavadati2013disfa} and FERA2015 \cite{FERA2015}), annotated in terms of AU intensity, to train our models for sequence modelling. This allows us to automatically obtain segments where target AUs are active (intensity $\neq$ 0) and non-active (intensity$=$0). 
Although this allowed us to narrow down the possible number of AU activations in target videos, the annotation process could not be fully automated. 
This is mainly because of a high number of false positives that can occur in such obtained automated annotations of AUs due to the training of target models being performed on videos from different datasets, which can in some instances differ significantly in lighting, head-pose and other conditions, from the SEWA videos. Once the automatic annotation is performed, several annotators have manually inspected the obtained active segments of target AUs, defining the starting and end frame of the AU within the segments classified as active by the model. 

    We used this annotations to train the AU detector of 5 facial action units (AU) from the basic SEWA dataset: inner eyebrow raiser (AU1, 109 examples), outer eyebrow raiser (AU2, 79 examples), eyebrow lowerer (AU4, 94 examples), lip corner puller (AU12, 104 examples), and chin raiser (AU17, 61 examples). 
    Similarity, the AU examples were again identified in a semi-automatic manner. Specifically, we first applied automatic AU detectors to the video segments and manually removed all false-positives from the detection results. Consequently, the AU annotation is not exhaustive, meaning that some AU activations may be missed. 

    Using this semi-automatic approach, we annotated 500 sequences (150 frames each) containing at least one of the 5 target AUs. We will refer to this dataset as \sewaAU. 
For the baseline experiments, we split this dataset in subject independent training, development and test sets. 
The size of each dataset and for each AU are shown in table \ref{tab:size}.

    The proposed method to semi-automatic AU detection has been implemented into a standalone module (VSL-AU detector) in C++/Matlab. This detector is then further integrated into the SEWA back-end emotion recognition server using the HCI2  Framework \cite{shen2013framework}. 

\begin{table}[htbp]
  \centering
  \caption{Number of frames with active AUs in training (\emph{TR}), test (\emph{TE}) and validation (\emph{VA}) set.}
    \begin{tabular*}{0.6\columnwidth}{@{\extracolsep{\fill}} lrrr}
    \toprule
    \textbf{AU} & \textbf{TR} & \textbf{TE} & \textbf{VA} \\
    \midrule
    \textbf{1} & 5180  & 4340  & 5740 \\
    \textbf{2} & 4060  & 3220  & 3920 \\
    \textbf{4} & 4620  & 4340  & 4200 \\
    \textbf{12} & 5600  & 4620  & 4340 \\
    \textbf{17} & 3500  & 3080  & 2940 \\
    \textbf{Total} & 22960 & 19600 & 21140 \\
    \bottomrule
    \end{tabular*}
  \label{tab:size}
\end{table}%
\subsubsection{Valence, Arousal, and Liking/Disliking annotation}
\begin{figure}
    \centering
    \includegraphics[width=1.\linewidth]{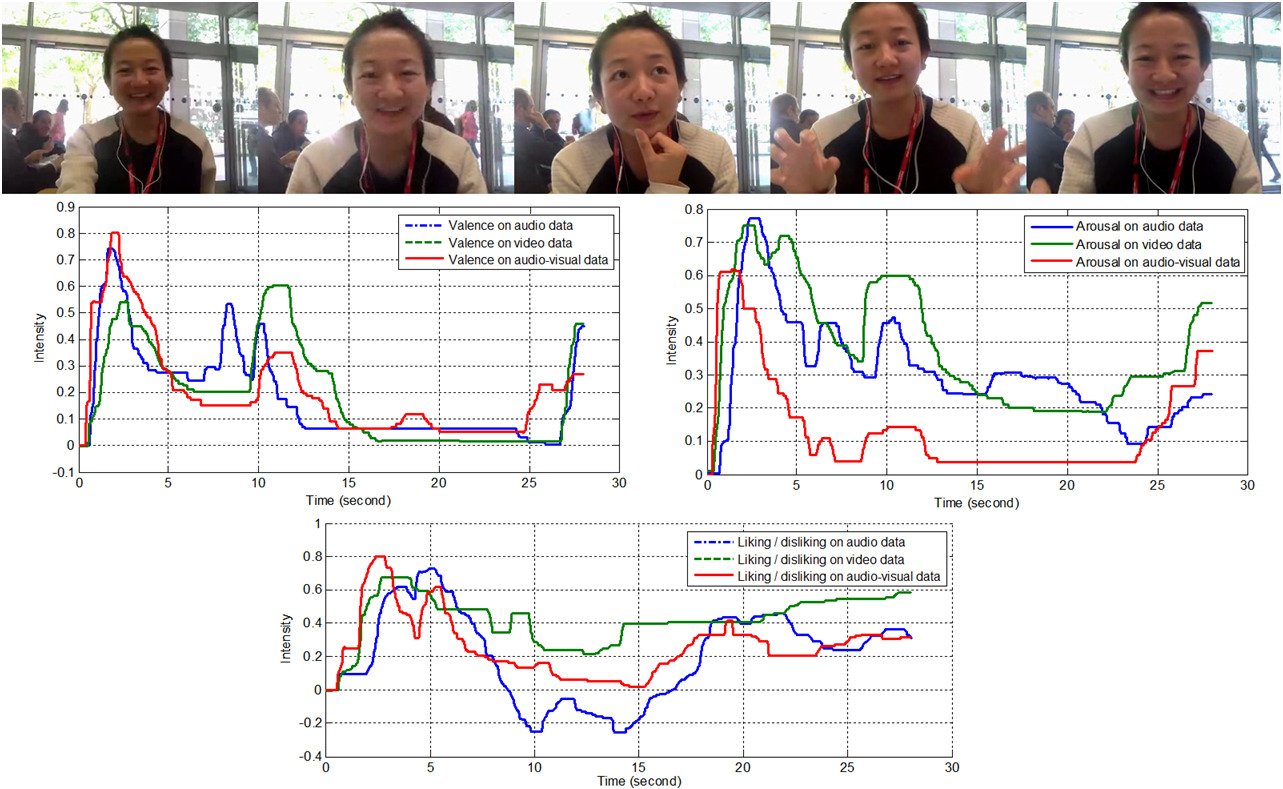}
    \caption{An example of the continuously valued annotation results on valence, arousal and liking and disliking.}
    \label{fig:valence_arousal_annotations}
\end{figure}

Continuously-valued valance, arousal and liking / disliking (toward the advertisement) were annotated for all segments in the basic SEWA dataset. In order to identify the subtle changes in the subjects’ emotional state, annotators were always hired from the same cultural background of the recorded subjects. In addition, to reduce the effect of the annotator bias, 5 annotators were recruited for each culture. The annotation was performed using a custom-built tool, which played the recordings while asked the annotators to push / pull a joystick in real-time to indicate the subject's level of valence, arousal, or liking/disliking. The joystick's pitch value was then sampled at approximately 66 Hz and saved as the annotation. To avoid cognitive overload on the annotators, the three dimensions (valence, arousal and liking / disliking) were annotated separately in three passes. Furthermore, for each dimension, the segments were annotated three times, first based on audio data only, then based on video data only and finally based on audio-visual data. An example of the continuous annotations obtained with this process is illustrated in Fig. \ref{fig:valence_arousal_annotations}. Only the segments where the subject on camera was speaking and his chat partner silent were considered. Of the segments satisfying this condition, 90 were selected for annotation so as to contain 15 segments for each of the following criteria: (a) high arousal, (b) low arousal, (c) positive valence, (d) negative valence, (e) presence of liking and (f) presence of disliking. The latter two 30 segments were continuously annotated for liking and disliking while all 90 segments were fully annotated for continuous valence and arousal. 

    These continuous annotations obtained are further combined into one single ground-truth employing Canonical Time Warping for each sequence to construct a subspace were the annotations of all raters are maximally correlated with each other and with the corresponding audio-visual features. The ground-truth annotation is then derived from the correlated subspace. More precisely, it is obtained by keeping only the coefficient corresponding to the first component each annotation. This is additionally normalised in the continuous range \([0, 1]\).

\subsubsection{Behaviour Templates}

\begin{figure}[t!]
    \centering
    \includegraphics[width=1.\linewidth]{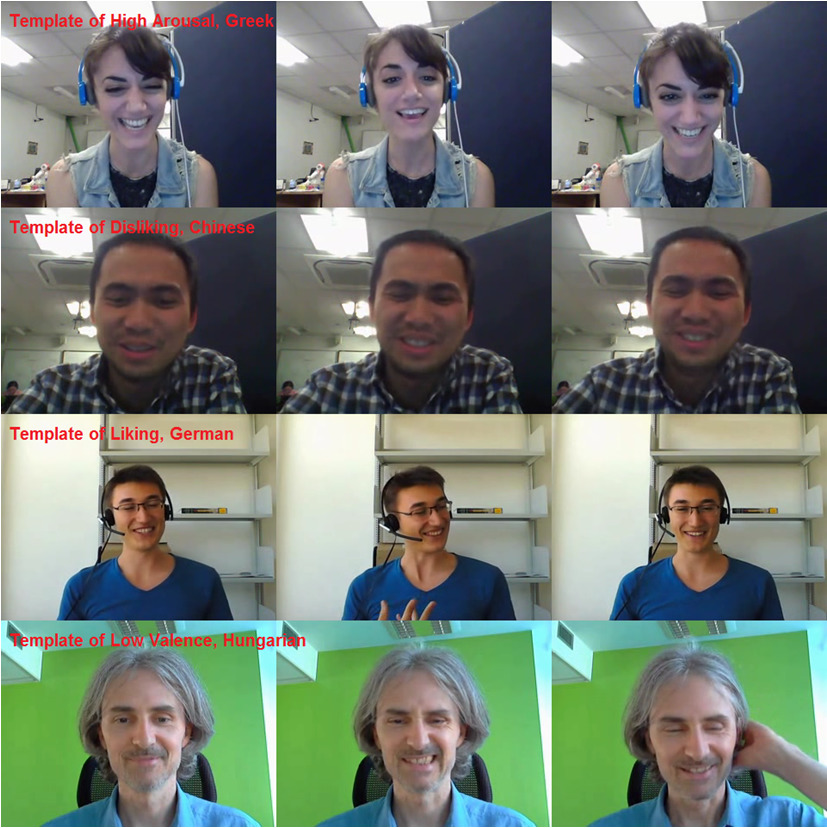}
        \caption{Examples of behaviour templates identified from the basic SEWA dataset.}
    \label{fig:behavior_templates}
\end{figure}

Moreover, we identified behaviour templates --that is prototypical behaviours-- for each culture when the subjects are in the emotional state of low / high valence, low / high arousal or showing liking / disliking toward the advertisement. For each category, at least two examples were identified. Table \ref{tab:templates} shows the exact distribution of the templates found in the basic SEWA dataset. These templates can be used to train and test the behaviour similarity detector. Figure.~\ref{fig:behavior_templates} illustrates some examples of these behaviour templates.

\begin{table}[]
\centering
\caption{Behaviour templates identified in the basic SEWA dataset.}
\label{tab:templates}
\begin{tabular*}{1\columnwidth}{@{\extracolsep{\fill}} lcccccc}
 		     \toprule
             & \textbf{British} & \textbf{German} & \textbf{Hungarian} & \textbf{Serbian} & \textbf{Greek} & \textbf{Chinese} \\ 
\midrule
\textbf{Low Valence}  & 2       & 4      & 2         & 6       & 2     & 3       \\ 
\textbf{High Valence} & 2       & 4      & 2         & 5       & 2     & 4       \\ 
\textbf{Low Arousal}  & 2       & 3      & 2         & 2       & 2     & 2       \\ 
\textbf{High Arousal} & 2       & 3      & 2         & 6       & 2     & 4       \\ 
\textbf{Liking}       & 2       & 4      & 2         & 6       & 2     & 5       \\ 
\bottomrule
\end{tabular*}
\end{table}
 
\begin{figure}
    \centering
    \begin{subfigure}{.44\linewidth} 
        \centering
        \includegraphics[width=0.95\linewidth]{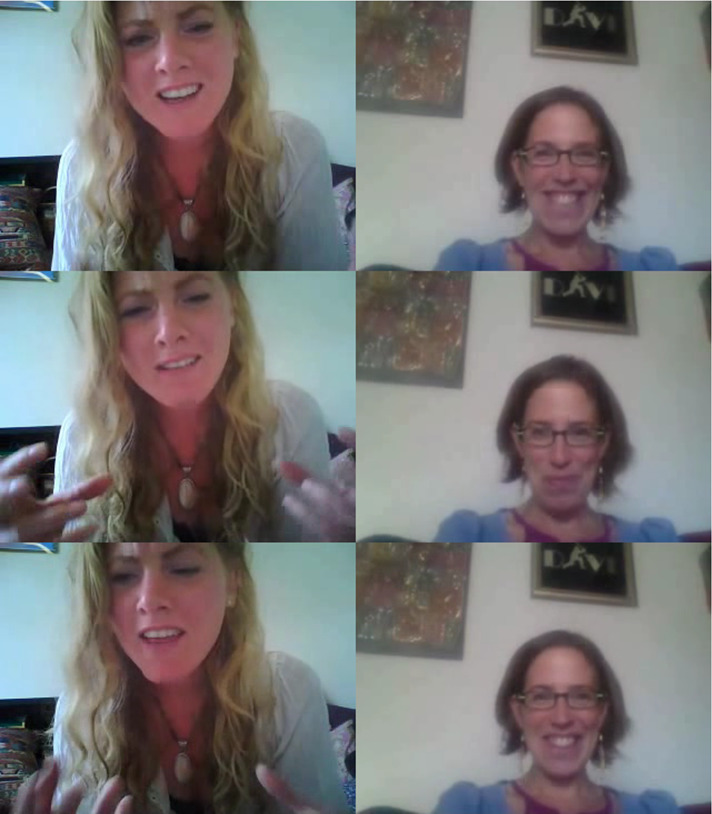}
        \caption{Strong agreement}
    \end{subfigure}
    \begin{subfigure}{.44\linewidth} 
        \centering
        \includegraphics[width=0.95\linewidth]{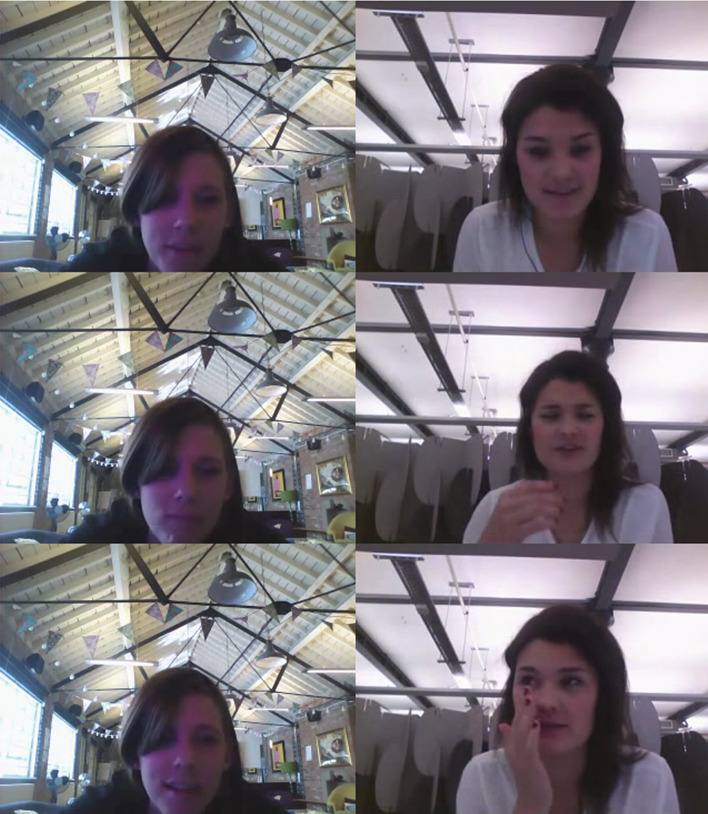}
        \caption{Strong disagreement}
    \end{subfigure}
    \caption{Examples of agreement and disagreement episodes.}
    \label{fig:agreement}
\end{figure}

In addition to continuous values such as valence and arousal, we extracted a number of episodes from the video-chat recordings in which the pair of subjects were in low, mid or high level of agreement / disagreement with each other and annotated the level of agreement/disagreement. The selections were based on the consensus of at least 3 annotators from the same culture of the recorded subjects. The exact numbers of agreement / disagreement episodes are shown in Table \ref{tab:agreement}. Two examples of the agreement / disagreement episodes are shown in Figure \ref{fig:agreement}.

\begin{table}[]
\centering
\caption{Agreement / disagreement episodes identified in the video-chat recordings.}
\label{tab:agreement}
\begin{tabular*}{1\columnwidth}{@{\extracolsep{\fill}} p{20mm}cccccc}
 \toprule
                               & \textbf{British} & \textbf{German} & \textbf{Hungarian} & \textbf{Serbian} & \textbf{Greek} & \textbf{Chinese} \\ \midrule
\textbf{Strong \newline Agreement}      & 12      & 7      & 7         & 7       & 5     & 5       \\ \hline
\textbf{Moderate \newline Agreement}    & 26      & 7      & 6         & 7       & 5     & 6       \\ \hline
\textbf{Weak \newline Agreement}       & 29      & 7      & 6         & 7       & 5     & 6       \\ \hline
\textbf{Weak \newline Disagreement}     & 7       & 6      & 5         & 4       & 5     & 4       \\ \hline
\textbf{Moderate \newline Disagreement} & 3       & 9      & 5         & 6       & 5     & 5       \\ \hline
\textbf{Strong \newline Disagreement}   & 3       & 6      & 5         & 4       & 5     & 3       \\ \bottomrule
\end{tabular*}
\end{table}
 
\subsubsection{Mimicry Episodes}
\begin{figure}
    \centering
    \begin{subfigure}{.44\linewidth} 
        \centering
        \includegraphics[width=0.95\linewidth]{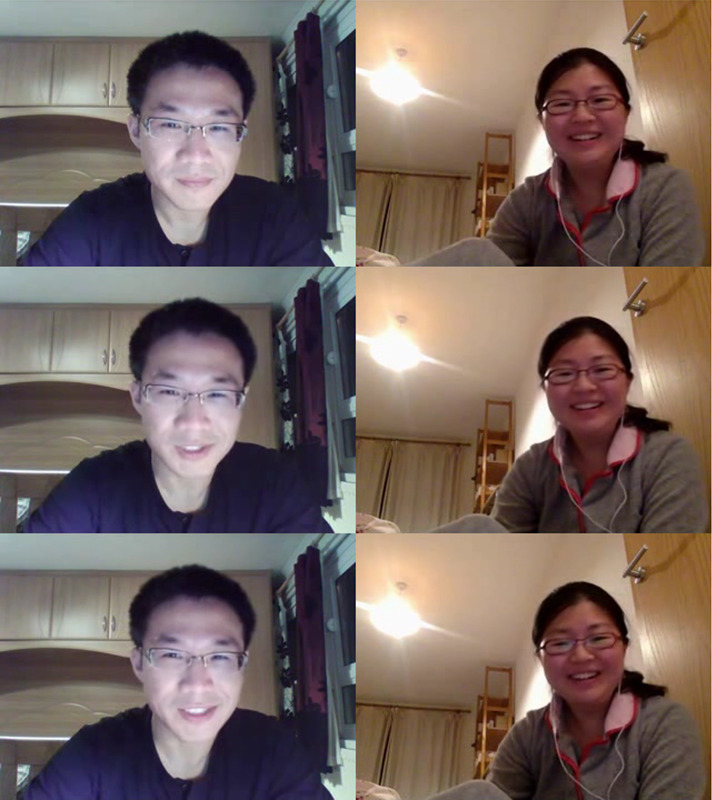}
        \caption{Chinese culture}
    \end{subfigure}
    \begin{subfigure}{.44\linewidth} 
        \centering
        \includegraphics[width=0.95\linewidth]{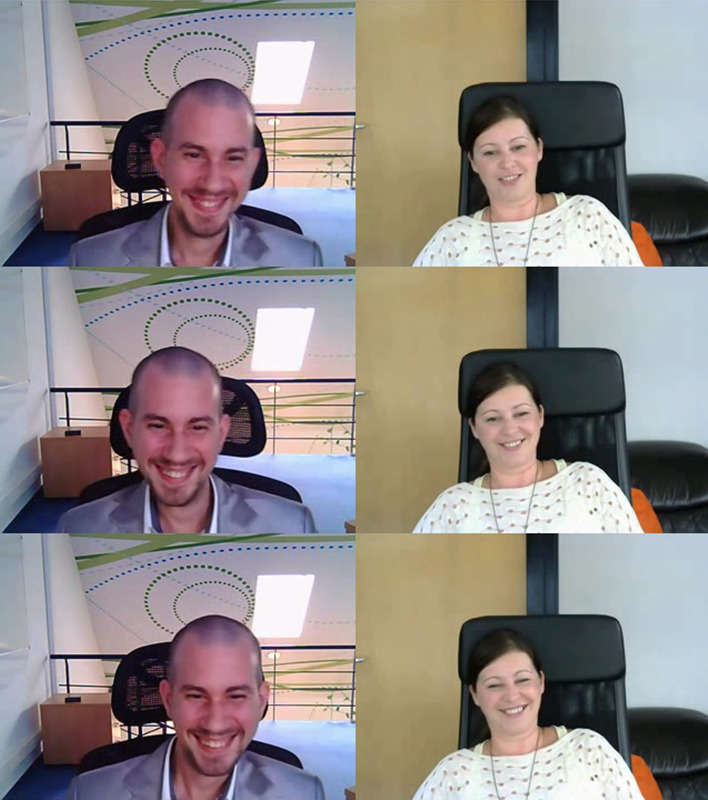}
        \caption{Hungarian culture}
    \end{subfigure}
    \caption{Examples of the mimicry episodes.}
    \label{fig:mimic}
\end{figure}

Lastly, 197 mimicry episodes (48 British, 31 German, 39 Hungarian, 20 Serbian, 41 Greek and 17 Chinese), in which one subject mimicked the facial expression and / or head gesture of the other subject, were identified from the video-chat recordings. Two examples of the identified mimicry episodes are shown in Figure \ref{fig:mimic}.

\subsection{Database availability}
The SEWA database is available online at: \url{http://db.sewaproject.eu/}. The web-portal provides a comprehensive search filter (shown in Fig. \label{fig:annotation_website}) allowing users to search for specific recordings based on various criteria, such as demographic data (gender, age, cultural background, etc.), availability of certain types of annotation, and so on. This will facilitate investigations during and beyond the project in the field of machine analysis of facial behaviour as well as in other research fields.

The SEWA database is made available to researchers for academic-use only. To comply with clauses stated in the Informed Consent signed by the recorded participants, all non-academic/commercial uses of the data are prohibited. Only researchers who signed the EULA will be granted access to the database. In order to ensure secure transfer of data from the database to an authorised user’s PC, the data are protected by SSL (Secure Sockets Layer) with an encryption key. If at any point, the administrators of the SEWA database and/or SEWA researchers have a reasonable doubt that an authorised user does not act in accordance to the signed EULA, they will be declined the access to the database. 
 \section{Baseline experiments}
\label{sec:exp}

In this section, we introduce the experimental setting and results for action unit detection, as well as valence, arousal and liking/disliking estimation.

\subsubsection{Methods}

 We performed experiments with three widely used and established methods:

\textbf{Support Vector Machine for Regression (SVR)}: We used a Support Vector Machine for regression, a common approach for affect estimation that has been widely used as a baseline for valence and arousal estimation \cite{avec_2012,avec_2014,avec_2016} and to produce state-of-the-art results \cite{baltrusaitis2013dimensional,nicolaou2012output}. 
In this paper, we use the Scikit-Learn implementation \cite{scikit-learn} and a linear kernel. \\

\textbf{Tree-based} The Random Forest Regressor (\emph{RF}) was used to produce the the second set of baseline results. It has been shown to produce state-of-the-art results on a wide range of problems \cite{evaluation_supervised_learning} and especially for continuous emotion recognition \cite{3D_emotion_random_forest}. The scikit-learn \cite{scikit-learn} implementation of random forests was used.\\

\textbf{Long Short Term Memory Recurrent Neural Networks (LSTM-RNN)} : We utilize LSTM-RNN as our third baseline method, owing to their popularity and their ability to learn long-range contextual information for sequential patterns~\cite{Hochreiter97-LST, Graves12-SSL} and their successful application for continuous emotion recognition~\cite{Zhang16-Facing, Han16-strength}.
To implement the LSTM-RNN models, we utilized the CURRENNT~\cite{Weninger15-ICT2} toolkit.\\

\textbf{Deep Convolutional Neural Networks (DCNN)}: We trained a ResNet-18 by optimizing either the RMSE or directly the CCC loss for both valence and arousal. We used a mini-batch size of 64. The network was trained end-to-end using Adam, with $\beta_1 = 0.5$ and $\beta_2 = 0.999$.
The learning rate was validated in a range $\{10, 0.00001\}$ and used a momentum of $0.9$, and a weight decay of $0.0001$. We decreased the learning rate by a factor of 10 every $15$ epochs. Since the deep neural networks take as input an image, we conducted the experiment with the annotations obtained using the video.

\subsection{Feature extraction}
\label{sec:features}
    In this section, we describe in detail the feature extraction procedure. For video features, we used appearance-based and geometric-based features. For audio features, low level descriptors (LLDs) were used. 

\subsubsection{Appearance features}
    To model appearance, we used dense SIFT \cite{sift_lowe}, which are much more robust than raw pixels. 
After facial landmarks have been detected, images are normalised in term of similarity transformation (translation, scaling and rotation). Dense SIFT features with 8-bins are then extracted from patches of size 11x11 around each of the facial landmarks. The resulting descriptors therefore encode both geometric and appearance features. We reduced the dimensionality of these feature vectors by applying Principle Component Analysis (PCA). In particular, we kept the 300 first component with the highest associated eigen-values to obtain a lower-dimensionality subspace on which we then project the appearance vectors to obtain a compact but informative facial representation.\\

\subsubsection{Geometric features}

    We also use geometric information directly obtained from the detected facial landmarks (shape features). After variations due to translation, scaling and in-plane rotation have been removed, the feature vector is then represented by the coordinates $[x_k, y_k]$ for $k \in \{1, ..., 49\}$ of the facial landmarks, stacked into a vector \((x_1, y_1, \cdots, x_{49}, y_{49}) \in \mathcal(R)^{98}\). 

    In particular, our shape normalization follows the approach \cite{bidirectional_aam,newton_aam,aam_tip} and leverages a linear shape model built from images annotated with $u = 49$ fiducial points. The annotated shapes are first normalized using Procrustes Analysis to remove variations due to similarity transformations (that is translation, rotation and scaling). From these we then obtain the aligned mean shape $\mathbf{s}_{0}$. To model similarity variations, we then explicitly construct 4 bases from $\mathbf{s}_{0}$ compactly represented as columns of $\mathbf{Q}^{2u \times 4}$. Given a shape $\mathbf{s}_t\in \mathcal{R}^{2u \times 1}$ a shape feature vector detected at frame \(t\), the similarity normalized features is then given by $\mathbf{s}_{\textrm{sim}} =  \mathbf{s}_y - \mathbf{Q}\mathbf{Q}^T(\mathbf{s}_y-\mathbf{s}_0)$.\\

\subsubsection{Audio features}

We used the established \cite{Eyben16-RSA} as our audio feature sets. For each audio recording, we capture the acoustic LLDs with the \textsc{openSMILE} toolkit~\cite{Eyben13-RDI} at a step size of $10\,ms$. Specifically, we extract frame-wise LLDs based on two different sets, namely, the {\em Interspeech 2013 Computational Para-linguistics Challenge (\textsc{ComParE}) set} and the {\em Geneva Minimalistic Acoustic Parameter Set (\textsc{GeMAPS})}, detailed descriptions of which will be given, respectively.

(\textsc{ComParE}) consists of 6\,373 acoustic features~\cite{Schuller13-TI2, Schuller14-TI2, Schuller15-TI2}. It contains 65 LLDs, covering spectral, cepstral, prosodic and voice quality information, which are summarised in Table~\ref{tab:compare}. From these LLDs extracted from each frame ($20\,ms\,-60\,ms$) of the audio signal, the first order derivatives (deltas) are computed and then functionals, such as, \eg moments and percentiles, are applied to each frame-level LLD and its delta coefficient over the whole audio signal, to form the \textsc{ComParE} feature set.  
In the feature sets provided with the SEWA database, however, deltas and functionals are not applied to enable the user to perform time-continuous emotion recognition. 
\subsubsection{Feature fusion}
Feature fusion is a method for combining different sets of features from different modalities to a resulting input vector with a combined descriptive power. The simplest form of feature fusion is early fusion which is done by concatenation of the different input vectors. More advanced methods like late fusion and mid-level fusion have show better performance but they can require special optimization and/or parameter tuning. In order to keep the experiments for this work simple and reproducible, we only use early fusion throughout all fusion based experiments.

\begin{table}[h]
\caption{INTERSPEECH 2013 Computational Paralinguistics Challenge feature set. Overview of 65 acoustic low-level descriptors (LLDs)}
 \begin{center}
\begin{tabular*}{1\columnwidth}{@{\extracolsep{\fill}} ll}
\toprule
\bf 4 energy related LLD & \bf Group\\
\midrule
Loudness & Prosodic\\
Modulation loudness & Prosodic\\
RMS energy, zero-crossing rate & Prosodic \\
\midrule
\bf 55 spectral related LLD&\bf  Group\\
\midrule
RASTA auditory bands 1-26 & Spectral \\
MFCC 1-14 & Cepstral \\
Spectral energy 250-650 Hz, 1-4 KHz & Spectral \\
Spectral roll-off Pt. .25, .50, .75, .90 & Spectral \\
Spectral flux, entropy, variance & Spectral \\
Spectral skewness and kurtosis & Spectral \\
Spectral slope & Spectral \\
Spectral harmonicity & Spectral \\
Spectral sharpness (auditory) & Spectral \\
Spectral centroid (linear) & Spectral \\
\midrule
\bf 6 voicing related LLD&\bf Group\\
\midrule
$F_0$ via SHS & Prosodic \\
Probability of voicing & Voice quality \\
Jitter (local and delta) & Voice quality \\
Shimmer & Voice quality \\
Log harmonics-to-noise ratio & Voice quality\\
\bottomrule
\end{tabular*}

\label{tab:compare}
\end{center}
\end{table}

The second acoustic feature set provided is based on \textsc{GeMAPS}~\cite{Eyben16-TGM}, a minimalistic expert-knowledge based feature set for the acoustic analysis of speaker states and traits. Compared with large-scale sets, such as \textsc{ComParE}, its main aim is to reduce the risk of over-fitting in the training phase. 
\textsc{GeMAPS} contains a compact set of 18 LLDs, covering spectral, prosodic and voice quality information, cf.\ Table~\ref{tab:gemaps}. The LLDs were selected with respect to their capability to describe affective physiological changes in voice production.

\begin{table}[h]
\caption{Geneva Minimalistic Acoustic Parameter Set. Overview of 18 acoustic low-level descriptors (LLDs)}
 \begin{center}
\begin{tabular*}{1\columnwidth}{@{\extracolsep{\fill}} ll}
\toprule
\bf 6 frequency related LLD& \bf Group\\
\midrule
Pitch & Prosodic\\
Jitter & Voice quality\\
Formant 1, 2, 3 frequency & Voice quality\\
Formant 1 bandwidth & Voice quality\\
\midrule

\bf 3 energy related LLD&\bf  Group\\
\midrule
Shimmer & Voice quality\\
Loudness & Prosodic\\
Harmonics-to-Noise ratio & Voice quality\\
\midrule
\bf 9 spectral related LLD&\bf Group\\
\midrule
$\alpha$ ratio & Spectral\\
Hammarberg Index & Spectral\\
Spectral slope 0-500 Hz and 500-1500Hz & Spectral\\
Formant 1, 2, 3 relative energy & Voice quality\\
Harmonic difference H1-H2, H1-A3  & Voice quality\\
\bottomrule
\end{tabular*}
\label{tab:gemaps}
\end{center}
\end{table}

For the baseline experiments, described next, the \textsc{ComParE} LLDs were summarized over a block of $6$ seconds computing the \textit{mean} and the \textit{standard deviation} of each LLD resulting in a feature vector of dimension 130. This is done as a single LLD frame does not convey meaningful information about the affective state of a speaker. Using \textsc{ComParE} only is justified by the fact that the LLDs in the \textsc{eGeMAPS} set are mostly redundant and the results achieved are not superior on average~\cite{Eyben16-TGM}.

\subsection{Experimental setting}

Extensive baseline experiments were conducted in several settings settings:

\textbf{Multi-culture, person independent experiment --coined \emph{multi}--}: This is the generic context in which we perform experiments on all cultures mixed (\ie training and testing on all cultures) but in a person-independent way.

\textbf{Culture independent --coined \( C1, \cdots, C6\)}: The goal of this setting is to test performance in a culture-specific manner (English, German, Hungarian, etc.). In this case, for each culture, the data of that culture was divided into person independent training, validation and testing sets.

In both cases, we ensured that the split of the data was person-independent by manually dividing the data into subject-independent training, development, and test partitions with a 3:1:1 ratio. All partitions were balanced with respect to age, gender and the criteria after that the segments have been selected, to make sure that we do not end up, e.\,g.\ in too many segments of liking in the training partition and in only segments of disliking in the test partition.

For each experiment, we optimised the model parameters by performing a grid-search on the development set to find the best setting of the regularization parameter $C$ for the SVR and number of trees $n$ for the Random Forest, and report results on the testing set.

\textbf{Deep based experiments}: in this case, we split the data in a person-independent manner by dividing the data into subject-independent training, development, and test partitions with a 8:1:1 ratio. We experimented with two different losses, either by optimizing the RMSE or by optimizing the CCC for both valence and arousal. The latter yields better results since it explicitly minimizes both the RMSE and the Correlation.

\subsubsection{Performance measure}

The problem of AU detection is a classification one while that of valence, arousal and liking/disliking level estimation is a regression one, mandating different error measures. Given a ground-truth and a prediction, for Action Units, we measure performance with the \(F_1\) score.  The \(f_1\) score is defined as:
\begin{equation}
    f_1 =  2 \cdot \frac{\mathrm{precision} \cdot \mathrm{recall}}{\mathrm{precision} + \mathrm{recall}}
\end{equation}

This score is widely used for AU-detection and classification of facial expressions of emotions \cite{walecki2015variable,automatic_survey_2015} because of its robustness to the imbalance in positive and negative samples, which is very common in the case of AUs.

For valence, arousal and liking/disliking, performance is measured using the Pearson product-moment correlation coefficient (\emph{CORR}), which is the standard measures used for measuring valence and arousal estimation accuracy \cite{gunes2013categorical}. We also report the Concordance Correlation Coefficient (\emph{CCC}), recently used in the last AVEC competitions \cite{Ringeval15-A2T,avec_2016}.

The correlation coefficient (CORR) is defined as follows. Let $\theta$ be a series of $n$ ground-truth labels, $n \in \mathcal{N}$ and $\hat\theta$ a series of $n$ corresponding prediction labels.

\begin{equation}
    \text{CORR}(\hat\theta, \theta) = 
    \frac{\text{COV} 
    ( \hat\theta, \theta)
    }{\sigma_{\hat\theta} 
    \sigma_{\theta}} = 
    \frac{E[(\hat\theta - 
    \mu_{\hat\theta})(\theta - \mu_{\theta})]}
    {\sigma_{\hat\theta}\sigma_{\theta}}
\end{equation}

Finally, the concordance correlation coefficient (CCC) is defined as:

\begin{equation}
	\text{CCC}(\hat\theta, \theta) =
    \frac{2 \times  \text{COV}( \hat\theta, \theta)}{
    	  \sigma_{\hat\theta}^2 + \sigma_{\theta}^2 + (\mu_{\hat\theta} - \mu_{\theta})^2},
    = \frac{2 E[(\hat\theta - \mu_{\hat\theta})(\theta - \mu_{\theta})]}{
    	  \sigma_{\hat\theta}^2 + \sigma_{\theta}^2 + (\mu_{\hat\theta} - \mu_{\theta})^2},
\end{equation}

\subsection{Experimental results}
Here we present the experimental results for action unit detection and valence, arousal and liking/disliking estimation. 

\subsubsection{Action unit detection}
We used the SVM and the Random Forest for AU detection using geometric and appearance features and feature fusion as described in section \ref{sec:features}.
The results in terms of F1-score for per-frame detection are shown in Table \ref{tab:au2_test} on the test set, and in Table \ref{tab:au2_develop} on the development set.
The tables show that the AU detector perform well, but clearly not good enough for AU detection in a fully automatic manner.
AU 12 has the highest F1-score (0.618) with feature fusion and SVM classifier. 
These results demonstrate again that it is important to use both types of features, texture and appearance, to achieve superior results.
In particular, and in line with previous research, the average results achieved by landmarks are higher than those by texture features which confirms the representative power of geometric features.
In comparison to the baseline results with those in the FERA2015 \cite{FERA2015} database, our results obtained here are lower on the overlapping AUs (10 and 17).
This is mainly because the SEWA dataset contains facial expressions recorded in different contexts and in the wild, while the FERA2015 recordings are made in an controlled environment or laboratory with controlled noise level, illumination and calibrated cameras. 
\begin{table}[htb]
  \centering
  \caption{F1-score for AU detection on the test partition}
    \begin{tabular*}{\columnwidth}{@{\extracolsep{\fill}} lrrrrrr}
    \toprule
          & \multicolumn{2}{c}{\textbf{Landmarks}} & \multicolumn{2}{c}{\textbf{SIFT}} & \multicolumn{2}{c}{\textbf{Fusion}} \\
          \cline{2-3}  \cline{4-5} \cline{6-7}
    \textbf{AU} & \textbf{SVM} & \textbf{RF} & \textbf{SVM} & \textbf{RF} & \textbf{SVM} & \textbf{RF} \\
    \midrule
    \textbf{1}  & 0.401 & 0.285 & 0.512 & 0.265 & 0.514 & 0.272 \\
    \textbf{2}  & 0.323 & 0.415 & 0.300 & 0.211 & 0.293 & 0.275 \\
    \textbf{4}  & 0.409 & 0.123 & 0.345 & 0.265 & 0.345 & 0.183 \\
    \textbf{12} & 0.513 & 0.492 & 0.518 & 0.321 & 0.613 & 0.421 \\
    \textbf{17} & 0.361 & 0.068 & 0.303 & 0.177 & 0.302 & 0.247 \\
    \textbf{av} & 0.385 & 0.251 & 0.378 & 0.226 & 0.407 & 0.271 \\
    \bottomrule
    \end{tabular*}
  \label{tab:au2_test}
\end{table}

\begin{table}[htb]
  \centering
  \caption{F1-score for AU detection on the development partition}
    \begin{tabular*}{\columnwidth}{@{\extracolsep{\fill}} lrrrrrr}
    \toprule
                & \multicolumn{2}{c}{\textbf{Landmarks}} & \multicolumn{2}{c}{\textbf{SIFT}} & \multicolumn{2}{c}{\textbf{Fusion}} \\
                \cline{2-3}  \cline{4-5} \cline{6-7}
    \textbf{AU} & \textbf{SVM}                            & \textbf{RF}                        & \textbf{SVM}                         & \textbf{RF} & \textbf{SVM} & \textbf{RF} \\
    \midrule
    \textbf{1}  & 0.345                                   & 0.198                              & 0.477                                & 0.161       & 0.479        & 0.301       \\
    \textbf{2}  & 0.583                                   & 0.470                              & 0.406                                & 0.276       & 0.404        & 0.271       \\
    \textbf{4}  & 0.405                                   & 0.255                              & 0.461                                & 0.290       & 0.460        & 0.289       \\
    \textbf{12} & 0.533                                   & 0.421                              & 0.588                                & 0.413       & 0.618        & 0.431       \\
    \textbf{17} & 0.419                                   & 0.282                              & 0.271                                & 0.297       & 0.271        & 0.246       \\
    \textbf{av} & 0.432                                   & 0.296                              & 0.417                                & 0.261       & 0.429        & 0.290       \\
    \bottomrule
    \end{tabular*}
  \label{tab:au2_develop}
\end{table}

\subsubsection{Estimation of valence, arousal and liking/disliking}

\begin{table}[]
    \centering
    \setlength\tabcolsep{0pt}
    \setlength\extrarowheight{2pt}
    \begin{tabular*}{1\columnwidth}{@{\extracolsep{\fill}} c c c c c c}
		\toprule
        \multicolumn{2}{c}{ } &    \multicolumn{2}{c}{\textbf{Valence}} & \multicolumn{2}{c}{\textbf{Arousal}}  \\ \cline{3-4} \cline{5-6}
        \textbf{Method} &  \textbf{Criterion}& \textbf{PCC} & \textbf{CCC} 
        &  \textbf{PCC} & \textbf{CCC} \\ \midrule
         & \textbf{RMSE} &  0.29 & 0.27 & 0.13 & 0.11 \\
        \cline{2-6}
        \multirow{-2}{*}{\textbf{ResNet}} & \textbf{CCC} & 0.35 & 0.35 & 0.35 & 0.29 \\
        \bottomrule
    \end{tabular*}
    \caption{Results for valence and arousal estimation using deep convolutional neural networks. The first row shows the results obtained with a ResNet-18, by optimizing the RMSE, while the second row was obtained by optimizing the CCC directly.}
    \label{tab:deep}
\end{table}

The setting of our baseline experiment allows us to investigate the effect of audio, video and the fusion of both on the results. In addition, we are able to separate the effect of culture on the results. As annotations were performed \emph{separately} but by the \emph{same annotators} on the audio, video and audio-video feeds respectively, we are also able to infer the human-level-performance of recognizing the levels of valence and arousal displayed by a subject given each type of information. Results are reported in term of CORR in Table. Table.~\ref{tab:results_pcc} and in term of CCC in Table.~\ref{tab:results_ccc}. In all cases, we report results when training with annotations and features obtained using either exclusively audio, video or audio-video.

\begin{table*}[]
    \centering
        \smallskip\noindent
        \resizebox{0.9\textwidth}{0.48\textheight}{ \npdecimalsign{.}
\nprounddigits{3}

\setlength\tabcolsep{0pt}
\setlength\extrarowheight{2pt}
\begin{tabular*}{\textwidth}{@{\extracolsep{\fill}} c  c  c  n{5}{3}  n{4}{3}  n{5}{3}  n{5}{3}  n{5}{3}  n{5}{3}  n{5}{3}  n{5}{3}  n{5}{3}  }
		\toprule
        \multicolumn{3}{c}{ } &    \multicolumn{3}{c}{\bf Valence} & \multicolumn{3}{c}{\bf Arousal}    & \multicolumn{3}{c}{\bf Liking/Disliking}         \\ 
        \cline{4-6} \cline{7-9} \cline{10-12}
\multicolumn{1}{c}{\bf Method} & \multicolumn{1}{c}{\bf Feature} & \multicolumn{1}{c}{\bf Annotation} & \multicolumn{1}{c}{\bf A }                 & \multicolumn{1}{c}{\bf V }                 & \multicolumn{1}{c}{\bf AV}                 & \multicolumn{1}{c}{\bf A }                 & \multicolumn{1}{c}{\bf V }                & \multicolumn{1}{c}{\bf AV}                & \multicolumn{1}{c}{\bf A }                 & \multicolumn{1}{c}{\bf V }                 & \multicolumn{1}{c}{\bf AV}                 \\ \toprule
\multirow{ 24}{*}{\bf SVM}    & \multirow{ 8}{*}{\bf A}       & \textbf{C1}          & 0.13237            & 0.16998            & -0.069623          & 0.28832            & 0.28245           & 0.1931            & 0.30104            & 0.0084772          & -0.072409          \\
       &         & \textbf{C2}          & 0.012938           & 0.3955             & 0.19255            & 0.61144            & -0.0015269        & 0.083069          & 0.10377            & 0.17767            & 0.22364            \\
       &         & \textbf{C3}          & 0.25123            & 0.0871             & 0.11559            & 0.61856            & 0.54495           & 0.69359           & -0.11921           & -0.099694          & 0.0021774          \\
       &         & \textbf{C4}          & -0.068858          & 0.14932            & -0.0022426         & 0.10079            & -0.222            & 0.043052          & 0.49584            & 0.13187            & 0.63952            \\
       &         & \textbf{C5}          & -0.10594           & 0.39777            & 0.10905            & 0.067404           & 0.24343           & 0.35161           & 0.27868            & 0.1406             & 0.045579           \\
       &         & \textbf{C6}          & 0.027154           & 0.20902            & 0.22976            & 0.48823            & 0.49366           & 0.40213           & -0.25307           & 0.18613            & 0.28738            \\
       &         & \textbf{multi}       & 0.12331            & 0.29708            & 0.19568            & 0.42672            & 0.35121           & 0.26284           & 0.14441            & 0.22837            & 0.22892            \\
       &         & \textbf{avg.}        & 0.053172           & 0.243681428571429  & 0.1101092          & 0.371637714285714  & 0.241739014285714 & 0.289913          & 0.135922857142857  & 0.110489028571429  & 0.193543914285714  \\ \cline{2-12}
       & \multirow{ 8}{*}{\bf V}       & \textbf{C1}          & 0.193              & 0.427              & 0.294              & 0.097              & 0.228             & 0.228             & 0.163              & 0.299              & 0.342              \\
       &         & \textbf{C2}          & 0.238              & 0.154              & 0.203              & 0.174              & 0.457             & 0.337             & 0.08               & -0.06              & 0.188              \\
       &         & \textbf{C3}          & -0.081             & 0.278              & 0.215              & 0.055              & 0.386             & 0.193             & 0.101              & 0.417              & 0.027              \\
       &         & \textbf{C4}          & 0.005              & 0.301              & 0.376              & 0.244              & 0.236             & 0.376             & 0.231              & 0.165              & 0.39               \\
       &         & \textbf{C5}          & 0.155              & 0.495              & 0.252              & 0.019              & 0.271             & 0.15              & 0.494              & 0.17               & 0.406              \\
       &         & \textbf{C6}          & 0.038              & 0.264              & 0.171              & 0.099              & 0.268             & 0.297             & 0.004              & 0.325              & -0.036             \\
       &         & \textbf{multi}       & 0.195              & 0.312              & 0.194              & 0.249              & 0.202             & 0.172             & 0.117              & 0.154              & 0.048              \\
       &         & \textbf{avg.}        & 0.106142857142857  & 0.318714285714286  & 0.243571428571429  & 0.133857142857143  & 0.292571428571429 & 0.250428571428571 & 0.17               & 0.21               & 0.195              \\ \cline{2-12}
       & \multirow{ 8}{*}{\bf AV}      & \textbf{C1}          & 0.268              & 0.445              & 0.305              & 0.116              & 0.224             & 0.255             & 0.188              & 0.436              & 0.268              \\
       &         & \textbf{C2}          & 0.215              & 0.184              & 0.264              & 0.166              & 0.501             & 0.405             & 0.076              & -0.096             & 0.231              \\
       &         & \textbf{C3}          & 0.057              & 0.359              & 0.284              & 0.1                & 0.448             & 0.296             & 0.195              & 0.374              & 0.112              \\
       &         & \textbf{C4}          & 0.063              & 0.282              & 0.32               & 0.198              & 0.183             & 0.179             & 0.055              & 0.254              & 0.341              \\
       &         & \textbf{C5}          & 0.188              & 0.468              & 0.236              & 0.095              & 0.261             & 0.217             & 0.452              & 0.165              & 0.406              \\
       &         & \textbf{C6}          & -0.02              & 0.296              & 0.212              & 0.035              & 0.229             & 0.19              & 0.242              & 0.294              & -0.05              \\
       &         & \textbf{multi}       & 0.171              & 0.326              & 0.199              & 0.267              & 0.164             & 0.175             & 0.103              & 0.148              & 0.054              \\
       &         & \textbf{avg.}        & 0.134571428571429  & 0.337142857142857  & 0.26               & 0.139571428571429  & 0.287142857142857 & 0.245285714285714 & 0.187285714285714  & 0.225              & 0.194571428571429  \\ \midrule
\multirow{ 24}{*}{\bf RF}     & \multirow{ 8}{*}{\bf A}       & \textbf{C1}          & 0.203              & 0.116              & 0.018              & 0.341              & 0.169             & 0.235             & 0.264              & -0.07              & 0.004              \\
       &         & \textbf{C2}          & 0.06               & 0.117              & -0.037             & -0.023             & 0.433             & -0.141            & -0.029             & 0.184              & 0.007              \\
       &         & \textbf{C3}          & 0.165              & 0.054              & 0.105              & 0.665              & 0.388             & 0.531             & -0.191             & -0.066             & 0.069              \\
       &         & \textbf{C4}          & -0.13              & -0.175             & -0.017             & 0.067              & 0.054             & 0.09              & 0.303              & -0.016             & 0.182              \\
       &         & \textbf{C5}          & 0.002              & 0.154              & 0.206              & 0.007              & 0.217             & 0.062             & 0.055              & 0.111              & 0.203              \\
       &         & \textbf{C6}          & 0.115              & 0.238              & 0.155              & 0.56               & 0.37              & 0.363             & 0.002              & 0.106              & 0.088              \\
       &         & \textbf{multi}       & 0.104              & 0.156              & 0.078              & 0.294              & 0.288             & 0.223             & 0.078              & 0.121              & 0.059              \\
       &         & \textbf{avg.}        & 0.0741428571428572 & 0.0942857142857143 & 0.0725714285714286 & 0.273              & 0.274142857142857 & 0.194714285714286 & 0.0688571428571429 & 0.0528571428571429 & 0.0874285714285714 \\ \cline{2-12}
       & \multirow{ 8}{*}{\bf V}       & \textbf{C1}          & 0.002              & 0.366              & 0.151              & 0.104              & 0.123             & 0.166             & 0.324              & 0.109              & 0.319              \\
       &         & \textbf{C2}          & 0.115              & -0.007             & 0.099              & 0.127              & 0.511             & 0.204             & 0.156              & -0.042             & -0.07              \\
       &         & \textbf{C3}          & 0.123              & 0.238              & 0.177              & 0.129              & 0.104             & 0.255             & -0.078             & 0.021              & 0.158              \\
       &         & \textbf{C4}          & 0.047              & 0.387              & 0.324              & 0.05               & 0.234             & 0.193             & 0.185              & 0.027              & 0.304              \\
       &         & \textbf{C5}          & 0.037              & 0.259              & 0.129              & -0.016             & 0.265             & 0.134             & 0.114              & -0.093             & -0.035             \\
       &         & \textbf{C6}          & 0.162              & 0.358              & 0.396              & 0.077              & 0.174             & 0.139             & 0.078              & 0.247              & 0.075              \\
       &         & \textbf{multi}       & 0.034              & 0.207              & 0.192              & 0.047              & 0.123             & 0.127             & -0.023             & 0.062              & 0.077              \\
       &         & \textbf{avg.}        & 0.0742857142857143 & 0.258285714285714  & 0.209714285714286  & 0.074              & 0.219142857142857 & 0.174             & 0.108              & 0.0472857142857143 & 0.118285714285714  \\ \cline{2-12}
       & \multirow{ 8}{*}{\bf AV}      & \textbf{C1}          & 0.135              & 0.358              & 0.091              & 0.027              & 0.225             & 0.056             & 0.347              & 0.293              & 0.376              \\
       &         & \textbf{C2}          & 0.064              & 0.126              & 0.133              & 0.145              & 0.355             & 0.158             & -0.064             & 0.136              & -0.081             \\
       &         & \textbf{C3}          & 0.064              & 0.193              & 0.226              & 0.056              & 0.12              & 0.115             & -0.029             & 0.177              & 0.083              \\
       &         & \textbf{C4}          & 0.058              & 0.374              & 0.256              & 0.084              & 0.152             & 0.162             & 0.13               & 0.045              & 0.347              \\
       &         & \textbf{C5}          & 0.043              & 0.262              & 0.119              & 0.052              & 0.241             & 0.079             & 0.109              & -0.133             & -0.121             \\
       &         & \textbf{C6}          & 0.103              & 0.36               & 0.335              & 0.028              & 0.173             & 0.089             & 0.078              & 0.272              & 0.032              \\
       &         & \textbf{multi}       & 0.023              & 0.22               & 0.137              & 0.043              & 0.143             & 0.125             & -0.065             & 0.113              & 0.071              \\
       &         & \textbf{avg.}        & 0.07               & 0.270428571428571  & 0.185285714285714  & 0.0621428571428571 & 0.201285714285714 & 0.112             & 0.0722857142857143 & 0.129              & 0.101              \\ \midrule
\multirow{ 24}{*}{\bf LSTM}   & \multirow{ 8}{*}{\bf A}       & \textbf{C1}          & 0.164937           & 0.134964           & 0.119517           & 0.346611           & 0.099328          & 0.320616          & 0.415297           & 0.206101           & 0.269366           \\
       &         & \textbf{C2}          & 0.317342           & 0.187504           & 0.215827           & 0.212382           & 0.07234           & 0.253195          & 0.221883           & 0.146281           & 0.163776           \\
       &         & \textbf{C3}          & 0.249168           & 0.1524             & 0.225657           & 0.66888            & 0.29783           & 0.539534          & 0.108054           & 0.118181           & 0.404093           \\
       &         & \textbf{C4}          & 0.131852           & 0.251299           & 0.300448           & 0.119773           & 0.110081          & 0.209711          & 0.172925           & 0.148906           & 0.294668           \\
       &         & \textbf{C5}          & 0.121331           & 0.380592           & 0.279451           & 0.115956           & 0.115429          & 0.238769          & 0.34887            & 0.324281           & 0.406523           \\
       &         & \textbf{C6}          & 0.238486           & 0.325534           & 0.304058           & 0.500465           & 0.532226          & 0.615697          & 0.219127           & 0.141321           & 0.086768           \\
       &         & \textbf{multi}       & 0.117972           & 0.212005           & 0.082011           & 0.346372           & 0.295699          & 0.233559          & 0.215              & 0.205717           & 0.151212           \\
       &         & \textbf{avg.}        & 0.191584           & 0.234899714285714  & 0.218138428571429  & 0.330062714285714  & 0.217561857142857 & 0.344440142857143 & 0.243022285714286  & 0.184398285714286  & 0.253772285714286  \\ \cline{2-12}
       & \multirow{ 8}{*}{\bf V}       & \textbf{C1}          & 0.07623            & 0.112386           & 0.074391           & 0.105743           & 0.180308          & 0.112096          & 0.19736            & 0.146239           & 0.242864           \\
       &         & \textbf{C2}          & 0.166543           & 0.083945           & 0.151074           & 0.221196           & 0.106176          & 0.124645          & 0.248034           & 0.052              & 0.144003           \\
       &         & \textbf{C3}          & 0.115779           & 0.167894           & 0.190153           & 0.170786           & 0.186661          & 0.240166          & 0.055308           & 0.251871           & 0.135562           \\
       &         & \textbf{C4}          & 0.109293           & 0.009792           & 0.08138            & 0.083338           & 0.119615          & 0.191503          & 0.169146           & 0.095735           & -0.022043          \\
       &         & \textbf{C5}          & 0.212188           & 0.325203           & 0.170874           & 0.224732           & 0.243132          & 0.254236          & 0.251265           & 0.255736           & 0.259102           \\
       &         & \textbf{C6}          & 0.29516            & 0.245305           & 0.158454           & 0.199296           & 0.136734          & 0.203938          & 0.106613           & 0.12854            & 0.241313           \\
       &         & \textbf{multi}       & 0.090659           & 0.281281           & 0.15322            & 0.114934           & 0.115415          & 0.11897           & 0.164133           & 0.076151           & 0.085979           \\
       &         & \textbf{avg.}        & 0.152264571428571  & 0.175115142857143  & 0.139935142857143  & 0.160003571428571  & 0.155434428571429 & 0.177936285714286 & 0.170265571428571  & 0.143753142857143  & 0.155254285714286  \\ \cline{2-12}
       & \multirow{ 8}{*}{\bf AV}      & \textbf{C1}          & 0.235836           & 0.237997           & 0.231102           & 0.243861           & 0.23845           & 0.194512          & 0.181382           & 0.114215           & 0.172098           \\
       &         & \textbf{C2}          & -0.007543          & 0.128308           & 0.067364           & 0.093176           & 0.172477          & 0.186679          & 0.003383           & 0.015112           & -0.000483          \\
       &         & \textbf{C3}          & 0.201745           & 0.159766           & 0.23217            & 0.168039           & 0.128243          & 0.128133          & -0.03619           & 0.119753           & 0.269981           \\
       &         & \textbf{C4}          & 0.106559           & 0.056096           & 0.138185           & 0.122388           & 0.141391          & 0.185742          & 0.094735           & 0.025932           & 0.054617           \\
       &         & \textbf{C5}          & 0.101418           & 0.241422           & 0.242896           & 0.110516           & 0.104805          & 0.250952          & 0.152956           & 0.178612           & 0.200518           \\
       &         & \textbf{C6}          & 0.093054           & 0.140216           & 0.253513           & 0.178226           & 0.221367          & 0.287486          & 0.233642           & 0.063548           & 0.119799           \\
       &         & \textbf{multi}       & 0.118838           & 0.227553           & 0.194602           & 0.166713           & 0.112282          & 0.162288          & 0.094521           & 0.064183           & 0.065073           \\
       &         & \textbf{avg.}        & 0.121415285714286  & 0.170194           & 0.194261714285714  & 0.154702714285714  & 0.159859285714286 & 0.199398857142857 & 0.103489857142857  & 0.0830507142857143 & 0.125943285714286  \\ \bottomrule
\end{tabular*}
  }
    \caption{Results in term of CCC for valence, arousal and liking/disliking}
    \label{tab:results_ccc}
\end{table*}

Results show that, for valence, we obtained better results on annotation obtained using exclusively video. These results are slightly lower when using labels obtained by annotating audio-video, while the worst results were obtained on the labels collected from the audio feed only. This is in line with the recent finding by psychologists that valence
is much better estimated from video imagery than from audio only,
while arousal is much better predicted from audio than from video \cite{russel2003,Kroschel07}. As expected, using a fusion of audio-video features increases the results, while audio features are the least helpful, supporting the theory that the face and its deformation is the main medium of communication between humans when it comes to emotions.

\begin{table*}[]
    \centering
        \smallskip\noindent
        \resizebox{0.9\textwidth}{0.48\textheight}{ \npdecimalsign{.}
\nprounddigits{3}

\setlength\tabcolsep{0pt}
\setlength\extrarowheight{2pt}
\begin{tabular*}{\textwidth}{@{\extracolsep{\fill}} c  c  c  n{5}{3}  n{4}{3}  n{5}{3}  n{5}{3}  n{5}{3}  n{5}{3}  n{5}{3}  n{5}{3}  n{5}{3}  }
		\toprule
        \multicolumn{3}{c}{ } &    \multicolumn{3}{c}{\bf Valence} & \multicolumn{3}{c}{\bf Arousal}    & \multicolumn{3}{c}{\bf Liking/Disliking}         \\ 
        \cline{4-6} \cline{7-9} \cline{10-12}
\multicolumn{1}{c}{\bf Method} & \multicolumn{1}{c}{\bf Feature} & \multicolumn{1}{c}{\bf Annotation} & \multicolumn{1}{c}{\bf A }                 & \multicolumn{1}{c}{\bf V }                 & \multicolumn{1}{c}{\bf AV}                 & \multicolumn{1}{c}{\bf A }                 & \multicolumn{1}{c}{\bf V }                & \multicolumn{1}{c}{\bf AV}                & \multicolumn{1}{c}{\bf A }                 & \multicolumn{1}{c}{\bf V }                 & \multicolumn{1}{c}{\bf AV}                 \\ \toprule
 \multirow{ 24}{*}{\bf SVM}           & \multirow{ 8}{*}{\bf A}       & \textbf{C1}          & 0.139             & 0.177             & -0.078             & 0.336              & 0.305             & 0.218             & 0.42               & -0.012               & -0.012             \\ 
               &         & \textbf{C2}          & -0.178            & 0.464             & -0.161             & 0.628              & -0.002            & 0.035             & 0.049              & 0.217                & -0.205             \\ 
               &         & \textbf{C3}          & 0.401             & 0.089             & 0.139              & 0.671              & 0.523             & 0.688             & 0.01               & 0.306                & 0.167              \\ 
               &         & \textbf{C4}          & -0.071            & -0.168            & -0.004             & 0.299              & -0.085            & 0.199             & 0.625              & 0.186                & 0.641              \\ 
               &         & \textbf{C5}          & -0.118            & 0.557             & 0.051              & 0.367              & 0.433             & 0.396             & 0.351              & 0.547                & 0.198              \\ 
               &         & \textbf{C6}          & -0.023            & 0.262             & 0.33               & 0.695              & 0.536             & 0.533             & 0.003              & 0.151                & 0.103              \\ 
               &         & \textbf{multi}       & 0.165             & 0.332             & 0.251              & 0.37               & 0.362             & 0.259             & 0.171              & 0.247                & 0.28               \\ 
               &         & \textbf{avg.}        & 0.045             & 0.244714285714286 & 0.0754285714285714 & 0.480857142857143  & 0.296             & 0.332571428571429 & 0.232714285714286  & 0.234571428571429    & 0.167428571428571  \\ \cline{2-12}
               & \multirow{ 8}{*}{\bf V}       & \textbf{C1}          & 0.197             & 0.433             & 0.347              & 0.097              & 0.272             & 0.266             & 0.201              & 0.333                & 0.397              \\ 
               &         & \textbf{C2}          & 0.231             & 0.189             & 0.222              & 0.235              & 0.46              & 0.444             & 0.051              & -0.069               & 0.179              \\ 
               &         & \textbf{C3}          & -0.059            & 0.334             & 0.229              & -0.008             & 0.443             & 0.229             & 0.117              & 0.542                & 0.034              \\ 
               &         & \textbf{C4}          & 0.182             & 0.403             & 0.463              & 0.338              & 0.349             & 0.254             & 0.14               & 0.236                & 0.432              \\ 
               &         & \textbf{C5}          & 0.167             & 0.47              & 0.301              & 0.022              & 0.289             & 0.22              & 0.57               & 0.208                & 0.408              \\ 
               &         & \textbf{C6}          & 0.173             & 0.389             & 0.319              & 0.102              & 0.348             & 0.235             & 0.001              & 0.431                & 0.027              \\ 
               &         & \textbf{multi}       & 0.171             & 0.321             & 0.266              & 0.281              & 0.182             & 0.209             & 0.132              & 0.134                & 0.024              \\ 
               &         & \textbf{avg.}        & 0.151714285714286 & 0.362714285714286 & 0.306714285714286  & 0.152428571428571  & 0.334714285714286 & 0.265285714285714 & 0.173142857142857  & 0.259285714285714    & 0.214428571428571  \\ \cline{2-12}
               & \multirow{ 8}{*}{\bf AV}     & \textbf{C1}          & 0.266             & 0.463             & 0.325              & 0.118              & 0.308             & 0.318             & 0.232              & 0.43                 & 0.477              \\ 
               &         & \textbf{C2}          & 0.247             & 0.109             & 0.239              & 0.222              & 0.423             & 0.351             & 0.081              & -0.107               & 0.16               \\ 
               &         & \textbf{C3}          & -0.137            & 0.407             & 0.317              & 0.13               & 0.437             & 0.273             & 0.229              & 0.467                & 0.122              \\ 
               &         & \textbf{C4}          & -0.064            & 0.392             & 0.434              & 0.327              & 0.355             & 0.325             & 0.206              & 0.319                & 0.383              \\ 
               &         & \textbf{C5}          & 0.186             & 0.489             & 0.37               & 0.097              & 0.258             & 0.29              & 0.5                & 0.185                & 0.432              \\ 
               &         & \textbf{C6}          & 0.062             & 0.27              & 0.12               & 0.032              & 0.28              & 0.231             & 0.102              & 0.411                & 0.05               \\ 
               &         & \textbf{multi}       & 0.203             & 0.333             & 0.268              & 0.282              & 0.171             & 0.169             & 0.192              & 0.157                & 0.057              \\ 
               &         & \textbf{avg.}        & 0.109             & 0.351857142857143 & 0.296142857142857  & 0.172571428571429  & 0.318857142857143 & 0.279571428571429 & 0.220285714285714  & 0.266                & 0.240142857142857  \\ \midrule
 \multirow{ 24}{*}{\bf RF}            & \multirow{ 8}{*}{\bf A}       & \textbf{C1}          & 0.289             & 0.162             & 0.025              & 0.387              & 0.242             & 0.334             & 0.519              & -0.107               & 0.008              \\ 
               &         & \textbf{C2}          & 0.156             & 0.134             & -0.112             & -0.044             & 0.562             & -0.384            & -0.06              & 0.385                & 0.013              \\ 
               &         & \textbf{C3}          & 0.328             & 0.079             & 0.166              & 0.772              & 0.492             & 0.647             & -0.424             & -0.214               & 0.102              \\ 
               &         & \textbf{C4}          & -0.18             & -0.324            & -0.087             & 0.135              & 0.123             & 0.18              & 0.523              & -0.022               & 0.454              \\ 
               &         & \textbf{C5}          & 0.006             & 0.502             & 0.378              & 0.03               & 0.472             & 0.139             & 0.199              & 0.375                & 0.662              \\ 
               &         & \textbf{C6}          & 0.171             & 0.341             & 0.287              & 0.682              & 0.556             & 0.573             & 0.003              & 0.156                & 0.154              \\ 
               &         & \textbf{multi}       & 0.227             & 0.268             & 0.14               & 0.399              & 0.411             & 0.312             & 0.168              & 0.198                & 0.116              \\ 
               &         & \textbf{avg.}        & 0.142428571428571 & 0.166             & 0.113857142857143  & 0.337285714285714  & 0.408285714285714 & 0.257285714285714 & 0.132571428571429  & 0.110142857142857    & 0.215571428571429  \\ \cline{2-12}
               & \multirow{ 8}{*}{\bf V}       & \textbf{C1}          & 0.139             & 0.392             & 0.266              & 0.144              & 0.138             & 0.094             & 0.346              & 0.09                 & 0.526              \\ 
               &         & \textbf{C2}          & 0.128             & 0.049             & 0.158              & 0.241              & 0.512             & 0.268             & 0.259              & -0.093               & -0.054             \\ 
               &         & \textbf{C3}          & 0.205             & 0.301             & 0.278              & 0.178              & 0.136             & 0.294             & 0.05               & -0.124               & 0.255              \\ 
               &         & \textbf{C4}          & 0.127             & 0.393             & 0.358              & 0.024              & 0.315             & 0.148             & 0.415              & -0.137               & 0.393              \\ 
               &         & \textbf{C5}          & 0.11              & 0.336             & 0.239              & -0.075             & 0.261             & 0.145             & 0.142              & -0.237               & 0.033              \\ 
               &         & \textbf{C6}          & 0.104             & 0.375             & 0.455              & 0.054              & 0.219             & 0.213             & 0.24               & 0.316                & 0.081              \\ 
               &         & \textbf{multi}       & 0.081             & 0.268             & 0.227              & 0.077              & 0.14              & 0.181             & 0.056              & 0.155                & 0.05               \\ 
               &         & \textbf{avg.}        & 0.127714285714286 & 0.302             & 0.283              & 0.0918571428571429 & 0.245857142857143 & 0.191857142857143 & 0.215428571428571  & -0.00428571428571428 & 0.183428571428571  \\ \cline{2-12}
               & \multirow{ 8}{*}{\bf AV}      & \textbf{C1}          & 0.226             & 0.457             & 0.09               & 0.059              & 0.251             & 0.173             & 0.261              & 0.356                & 0.505              \\ 
               &         & \textbf{C2}          & 0.107             & 0.188             & 0.212              & 0.211              & 0.46              & 0.222             & -0.111             & 0.079                & 0.062              \\ 
               &         & \textbf{C3}          & 0.098             & 0.303             & 0.233              & 0.175              & 0.157             & 0.191             & -0.124             & 0.227                & 0.055              \\ 
               &         & \textbf{C4}          & 0.042             & 0.341             & 0.353              & 0.137              & 0.219             & 0.183             & 0.239              & 0.053                & 0.475              \\ 
               &         & \textbf{C5}          & 0.162             & 0.318             & 0.177              & -0.029             & 0.399             & 0.218             & 0.121              & 0.123                & -0.182             \\ 
               &         & \textbf{C6}          & 0.11              & 0.416             & 0.38               & 0.102              & 0.223             & 0.149             & 0.144              & 0.343                & 0.128              \\ 
               &         & \textbf{multi}       & 0.028             & 0.243             & 0.201              & 0.097              & 0.164             & 0.197             & 0.04               & 0.147                & 0.113              \\ 
               &         & \textbf{avg.}        & 0.110428571428571 & 0.323714285714286 & 0.235142857142857  & 0.107428571428571  & 0.267571428571428 & 0.190428571428571 & 0.0814285714285714 & 0.189714285714286    & 0.165142857142857  \\ \midrule
 \multirow{ 24}{*}{\bf LSTM}          & \multirow{ 8}{*}{\bf A}       & \textbf{C1}          & 0.197294141185    & 0.144262368636    & 0.124715521998     & 0.389962834454     & 0.115240627505    & 0.345872095232    & 0.51849548346      & 0.221093375109       & 0.295517365017     \\ 
               &         & \textbf{C2}          & 0.331411157418    & 0.377160334989    & 0.34758319387      & 0.345333693537     & 0.115366814995    & 0.462364795823    & 0.452936995969     & 0.34674031124        & 0.385426425658     \\ 
               &         & \textbf{C3}          & 0.305601694368    & 0.158562609989    & 0.25772124517      & 0.680190630674     & 0.304326709421    & 0.568978505948    & 0.29923943314      & 0.189764193956       & 0.560170856275     \\ 
               &         & \textbf{C4}          & 0.136270066869    & 0.318239002587    & 0.353944570784     & 0.162042015413     & 0.134623532758    & 0.275507670419    & 0.220367921195     & 0.206096909105       & 0.350010502685     \\ 
               &         & \textbf{C5}          & 0.141533376136    & 0.524550136455    & 0.28808051148      & 0.226605122599     & 0.158286991017    & 0.281547669189    & 0.462985579723     & 0.492374803133       & 0.501132870325     \\ 
               &         & \textbf{C6}          & 0.327519075731    & 0.364519841895    & 0.306792129092     & 0.649086703401     & 0.66113110608     & 0.672102832518    & 0.303048192126     & 0.158110234228       & 0.1231190578       \\ 
               &         & \textbf{multi}       & 0.154             & 0.221             & 0.101              & 0.4                & 0.346             & 0.285             & 0.243              & 0.225                & 0.2                \\ 
               &         & \textbf{avg.}        & 0.227661358815286 & 0.301184899221571 & 0.254262453199143  & 0.407603000011143  & 0.262139397396571 & 0.413053367018429 & 0.357153372230429  & 0.262739975253       & 0.345053868251429  \\ \cline{2-12}
               & \multirow{ 8}{*}{\bf V}       & \textbf{C1}          & 0.104781588096    & 0.13691384463     & 0.0936034756266    & 0.137562537125     & 0.235718788532    & 0.139315333858    & 0.238210273246     & 0.152907249208       & 0.260340532461     \\ 
               &         & \textbf{C2}          & 0.403285996872    & 0.118992007867    & 0.273729933998     & 0.259436076228     & 0.115222649875    & 0.162020941779    & 0.388995063906     & 0.107221550272       & 0.467135641245     \\ 
               &         & \textbf{C3}          & 0.142638805227    & 0.188197225115    & 0.226836355717     & 0.207658862066     & 0.19025410777     & 0.307502706339    & 0.0970899825216    & 0.258181743491       & 0.218171431243     \\ 
               &         & \textbf{C4}          & 0.155032855481    & 0.0127492422399   & 0.103468541666     & 0.124704324547     & 0.184555424991    & 0.270596244878    & 0.227842432238     & 0.134621814653       & -0.0351408286006   \\ 
               &         & \textbf{C5}          & 0.235487166796    & 0.384219854327    & 0.187917111588     & 0.23736952761      & 0.269189527047    & 0.383028572874    & 0.25428554859      & 0.484737596358       & 0.261014497268     \\ 
               &         & \textbf{C6}          & 0.337641001928    & 0.310253357448    & 0.194553608277     & 0.320823311448     & 0.149889238682    & 0.274765996223    & 0.139350801498     & 0.144731115831       & 0.284699633161     \\ 
               &         & \textbf{multi}       & 0.135             & 0.322             & 0.187              & 0.14               & 0.148             & 0.173             & 0.186              & 0.11                 & 0.105              \\ 
               &         & \textbf{avg.}        & 0.216266773485714 & 0.2104750759467   & 0.181015575267514  & 0.203936377003428  & 0.184689962413857 & 0.244318542278714 & 0.218824871714229  & 0.198914438544714    & 0.223031558111057  \\ \cline{2-12}
               & \multirow{ 8}{*}{\bf AV}      & \textbf{C1}          & 0.328985816827    & 0.293320555657    & 0.285673594891     & 0.250371944676     & 0.330209187512    & 0.238735615143    & 0.194777714621     & 0.132061297934       & 0.178048796344     \\ 
               &         & \textbf{C2}          & -0.01404690269    & 0.133828281891    & 0.158452283208     & 0.104210709827     & 0.193453929999    & 0.251203689787    & 0.00686977988244   & 0.0357782157218      & -0.000731415139204 \\ 
               &         & \textbf{C3}          & 0.207057204412    & 0.218419099599    & 0.266487199186     & 0.200293184608     & 0.135853275492    & 0.147070739798    & -0.0380999125694   & 0.128593016487       & 0.271795091996     \\ 
               &         & \textbf{C4}          & 0.143871439411    & 0.0697311252648   & 0.189236243122     & 0.158000121661     & 0.174679441452    & 0.222883953304    & 0.113103188046     & 0.0291746032195      & 0.0897632003681    \\ 
               &         & \textbf{C5}          & 0.157421170528    & 0.285904644688    & 0.272684580938     & 0.138177269828     & 0.113893366498    & 0.302501871632    & 0.204757152357     & 0.276893720085       & 0.265547493537     \\ 
               &         & \textbf{C6}          & 0.0932933209524   & 0.202451821822    & 0.253914199977     & 0.268354053941     & 0.256978814672    & 0.351964726476    & 0.256785730005     & 0.0639303647617      & 0.132471451856     \\ 
               &         & \textbf{multi}       & 0.15              & 0.252             & 0.232              & 0.187              & 0.135             & 0.202             & 0.127              & 0.079                & 0.099              \\ 
               &         & \textbf{avg.}        & 0.152368864205771 & 0.207950789845971 & 0.236921157331714  & 0.186629612077286  & 0.191438287946429 & 0.245194370877143 & 0.12359909319172   & 0.106490174029857    & 0.147984945565985  \\ \bottomrule
\end{tabular*}
  }
    \caption{Results in term of CORR for valence, arousal and liking/disliking}
    \label{tab:results_pcc}
\end{table*}

On a model level, we observe that performances of different regression models vary from each other and that, overall, SVM performs better than RF which in turn outperforms LSTM. However, perhaps surprisingly, in the experiments on audio features, while Support Vector Machine for Regression (SVR) and Random Forest (RF) are expected to perform well for arousal and valence prediction, LSTM-RNNs noticeably outperform them for liking/disliking prediction. For example, while the average CCC of liking prediction based on audio features and A+V annotations is 0.194 and 0.087 by SVR and RF, respectively, a CCC of 0.254 is achieved with LSTM.

Still when it comes to audio features, in most cases, arousal is
better predicted than valence, which conforms repeated findings in the
literature \cite{nicolaou2012output,baltrusaitis2013dimensional,avec_2016}. Conversely, when using video features, valence seems to be more accurately predicted than arousal. These observation would confirm that acoustic features are more informative for arousal while valence can result in more subtle facial expressions requiring geometric and appearance features to be predicted accurately.
However, for liking or disliking, there is no such noticeable tendency. In most cases, as can be seen from the table, the performance for liking or disliking is lower than for arousal and valence. This could be mainly because the prediction of liking and disliking is more content-related and could not obtain sufficient useful information via acoustic cues only, lacking linguistic cues.

Moreover, regarding the three different types of annotations, we also note that, in most cases the best performance in terms of CCC was obtained by audio-based annotations for arousal and by video-based ones for valence, respectively, while no obvious performance improvement was seen when the combination of audio and video was provided during annotations. However, for liking or disliking, in many cases the best results of prediction of liking were achieved when the audio/video-based annotations were utilised. This may be because prediction of liking or disliking is a quite complex problem which is difficult to address with limited data. It could be improved when more data with information of multiple modalities is given.

Using video features, culture 5 (Hungarian) is best predicted (with a CCC of 0.495) for valence using SVR. Interestingly, this same culture is also best predicted for valence using audio features (CCC 0.398), again with SVR based on audio features but video-based annotation. For arousal and using audio features only, regarding the six different cultures, the performance in term of CCC (0.694) is obtained for culture 3 (German) with SVR on audio/video-based annotations. In contrast, using a fusion of audio \emph{and} video features, best results are obtained for arousal on culture 2 with a CCC of 0.501.

It is interesting to notice that, considering experiments of SVR, predictions of valence with video-based annotations outperforms that with audio-based annotations for all cultures except for culture 3 (German). Similarly, predictions of arousal with audio-based labels outperforms that with video-based labels for all cultures except for culture 5 (Hungarian) and 6 (Serbian). Such a contrast could be mainly due to the close connection between the two dimensions in spontaneous conversation. Therefore, it might be good to predict them together, \ie conducting multi-task learning to take advantage of the interconnections between the two different aspects.

Finally, as expected, experimental results indicate that ReseNet-18 (Table~\ref{tab:deep}) outperforms the other baseline models, particularly when trained by optimizing the CCC directly. This is especially true in the case of arousal, which is captured more accurately by deeply learned representations than by hand-crafted features. Valence on the other hand, as expected (e.g. \cite{AFEWVA_kossaifi}), can be predicted accurately from geometric features, encoded by facial landmarks or SIFT features.
However, this method is much more costly to run, both in terms of time and computational resources. \section{Conclusion}
    We introduced the SEWA database (SEWA DB), a multilingual dataset of annotated facial, vocal and verbal behaviour recordings made in-the-wild. In addition to providing training data for the technologies developed during the SEWA projects, the SEWA DB has also been made publicly available to the research community, representing a benchmark for efforts in automatic analysis of audio-visual behaviour in the wild. The SEWA DB contains the recordings of 204 experiment sessions, covering 408 subjects recruited from 6 different cultural backgrounds: British, German, Hungarian, Greek, Serbian, and Chinese. The database includes a total of 1525 minutes of audio-visual recordings of the subjects’ reaction to the 4 advertisement stimuli and 568 minutes of video-chat recordings of the subjects discussing the advertisement. In addition to the raw audio and video data, the SEWA DB also contains a wide range of annotations including: low-level audio descriptor (LLD) features, facial landmark locations, hand-gesture, head gesture, facial action units, audio transcript, continuously-valued valence, arousal and liking / disliking (toward the advertisement), template behaviours, agreement / disagreement episodes, and mimicry episodes.

	We provide exhaustive baseline experiments to assess Action Unit detection and valence, arousal and liking/disliking prediction, which is both helpful in advancing the field of affect estimation and will help advance the state-of-the art by providing a comparison benchmark.

    We believe this large corpus will be helpful to the community, both in the psychological field in helping test hypothesis and in the computer science field to advance the state of automatic sentiment analysis in the wild.

\ifCLASSOPTIONcompsoc
  \section*{Acknowledgments}
\else
  \section*{Acknowledgment}
\fi

This work was funded by the European Community Horizon 2020 [H2020/2014-
2020] under grant agreement no. 645094 (SEWA). 
Yannis Panagakis gratefully acknowledges the support of NVIDIA Corporation with the donation of the Titan Xp used for this research.

\ifCLASSOPTIONcaptionsoff
  \newpage
\fi

\bibliographystyle{IEEEtran}

\begin{thebibliography}{10}
\providecommand{\url}[1]{#1}
\csname url@samestyle\endcsname
\providecommand{\newblock}{\relax}
\providecommand{\bibinfo}[2]{#2}
\providecommand{\BIBentrySTDinterwordspacing}{\spaceskip=0pt\relax}
\providecommand{\BIBentryALTinterwordstretchfactor}{4}
\providecommand{\BIBentryALTinterwordspacing}{\spaceskip=\fontdimen2\font plus
\BIBentryALTinterwordstretchfactor\fontdimen3\font minus
  \fontdimen4\font\relax}
\providecommand{\BIBforeignlanguage}[2]{{
\expandafter\ifx\csname l@#1\endcsname\relax
\typeout{** WARNING: IEEEtran.bst: No hyphenation pattern has been}
\typeout{** loaded for the language `#1'. Using the pattern for}
\typeout{** the default language instead.}
\else
\language=\csname l@#1\endcsname
\fi
#2}}
\providecommand{\BIBdecl}{\relax}
\BIBdecl

\bibitem{brave2002emotion}
S.~Brave and C.~Nass, ``The human-computer interaction handbook,'' J.~A. Jacko
  and A.~Sears, Eds.\hskip 1em plus 0.5em minus 0.4em\relax Hillsdale, NJ, USA:
  L. Erlbaum Associates Inc., 2003, ch. Emotion in Human-computer Interaction,
  pp. 81--96.

\bibitem{pantic2003toward}
M.~Pantic and L.~J. Rothkrantz, ``Toward an affect-sensitive multimodal
  human-computer interaction,'' \emph{Proceedings of the IEEE}, vol.~91, no.~9,
  pp. 1370--1390, 2003.

\bibitem{d2015review}
S.~K. D'mello and J.~Kory, ``A review and meta-analysis of multimodal affect
  detection systems,'' \emph{ACM Computing Surveys (CSUR)}, vol.~47, no.~3,
  p.~43, 2015.

\bibitem{rinn1984neuropsychology}
W.~E. Rinn, ``The neuropsychology of facial expression: A review of the
  neurological and psychological mechanisms for producing facial expressions,''
  \emph{Psychological bulletin}, vol.~95, no.~1, pp. 52--77, 01 1984.

\bibitem{ambadar2009smiles}
\BIBentryALTinterwordspacing
Z.~Ambadar, J.~F. Cohn, and L.~I. Reed, ``All smiles are not created equal:
  Morphology and timing of smiles perceived as amused, polite, and
  embarrassed/nervous,'' \emph{Journal of nonverbal behavior}, vol.~33, no.~1,
  p. 17—34, March 2009. [Online]. Available:
  \url{http://europepmc.org/articles/PMC2701206}
\BIBentrySTDinterwordspacing

\bibitem{grimm2008vera}
M.~Grimm, K.~Kroschel, and S.~Narayanan, ``The vera am mittag german
  audio-visual emotional speech database,'' in \emph{ICME}.\hskip 1em plus
  0.5em minus 0.4em\relax IEEE, 2008, pp. 865--868.

\bibitem{mckeown2012semaine}
G.~McKeown, M.~Valstar, R.~Cowie, M.~Pantic, and M.~Schroder, ``{The SEMAINE
  database: Annotated multimodal records of emotionally colored conversations
  between a person and a limited agent},'' \emph{IEEE Transactions on Affective
  Computing}, vol.~3, no.~1, pp. 5--17, 2012.

\bibitem{CONFER}
C.~Georgakis, Y.~Panagakis, S.~Zafeiriou, and M.~Pantic, ``The conflict
  escalation resolution (confer) database,'' \emph{Image and Vision Computing},
  2017.

\bibitem{bilakhia2015mahnob}
S.~Bilakhia, S.~Petridis, A.~Nijholt, and M.~Pantic, ``The mahnob mimicry
  database: A database of naturalistic human interactions,'' \emph{Pattern
  recognition letters}, vol.~66, pp. 52--61, 2015.

\bibitem{Banziger12-ITG}
T.~B\"anziger, M.~Mortillaro, and K.~R. Scherer, ``{Introducing the Geneva
  Multimodal Expression Corpus for Experimental Research on Emotion
  Perception},'' \emph{Emotion}, vol.~12, no.~2, pp. 1161--1179, 2012.

\bibitem{DISFA}
S.~M. Mavadati, M.~H. Mahoor, K.~Bartlett, P.~Trinh, and J.~F. Cohn, ``Disfa: A
  spontaneous facial action intensity database,'' \emph{IEEE Trans. Affect.
  Comput.}, vol.~4, no.~2, pp. 151--160, Apr. 2013.

\bibitem{Koelstra12-TGM}
S.~Koelstra, C.~Muhl, M.~Soleymani, J.-S. Lee, A.~Yazdani, T.~Ebrahimi, T.~Pun,
  A.~Nijohlt, and I.~Patras, ``Deap: A database for emotion analysis using
  physiological signals,'' \emph{IEEE Transactions on Affective Computing},
  vol.~3, no.~1, pp. 18--31, 2012.

\bibitem{ringeval2013introducing}
F.~Ringeval, A.~Sonderegger, J.~Sauer, and D.~Lalanne, ``{{Introducing the
  RECOLA multimodal corpus of remote collaborative and affective
  interactions}},'' in \emph{Automatic Face and Gesture Recognition (FG), 2013
  10th IEEE International Conference and Workshops on}.\hskip 1em plus 0.5em
  minus 0.4em\relax IEEE, 2013, pp. 1--8.

\bibitem{Ekman11-WIM}
P.~Ekman and D.~Cordaro, ``What is meant by calling emotions basic,''
  \emph{Emotion Review}, vol.~3, no.~4, pp. 364--370, 2011.

\bibitem{Ekman93-FEA}
P.~Ekman, ``Facial expression and emotion,'' \emph{American Psychologist},
  vol.~28, no.~4, pp. 384--392, 1993.

\bibitem{Gross10-MP}
R.~Gross, I.~Matthews, J.~Cohn, T.~Kanade, and S.~Baker, ``Multi-pie,''
  \emph{Image and Vision Computing}, vol.~28, no.~5, pp. 807--813, 2010.

\bibitem{Busso08-IIE}
C.~Busso, M.~Bulut, C.-C. Lee, A.~Kazemzadeh, E.~Mower, S.~Kim, J.~N. Chang,
  S.~Lee, and S.~S. Narayanan, ``Iemocap: Interactive emotional dyadic motion
  capture database,'' \emph{Language resources and evaluation}, vol.~42, no.~4,
  pp. 335--359, 2008.

\bibitem{Zeng09-ASO}
Z.~Zeng, M.~Pantic, G.~I. Roisman, and T.~S. Huang, ``{A Survey of Affect
  Recognition Methods: Audio, Visual, and Spontaneous Expressions},''
  \emph{IEEE Transactions on Pattern Analysis and Machine Intelligence},
  vol.~31, no.~1, pp. 39--58, 2009.

\bibitem{gunes2013categorical}
H.~Gunes and B.~Schuller, ``Categorical and dimensional affect analysis in
  continuous input: Current trends and future directions,'' \emph{Image and
  Vision Computing}, vol.~31, no.~2, pp. 120--136, 2013.

\bibitem{Ayadi11-SOS}
M.~E. Ayadi, M.~S. Kamel, and F.~Karray, ``Survey on speech emotion
  recognition: Features, classification schemes, and databases,'' \emph{Pattern
  Recognition}, vol.~44, no.~3, pp. 572--587, 2011.

\bibitem{pantic2000expert}
M.~Pantic and L.~J. Rothkrantz, ``Expert system for automatic analysis of
  facial expressions,'' \emph{Image and Vision Computing}, vol.~18, no.~11, pp.
  881--905, 2000.

\bibitem{Bone14-RUA}
D.~Bone, C.-C. Lee, and S.~Narayanan, ``{Robust unsupervised arousal rating: A
  Rule-Based framework with knowledge-inspired vocal features},'' \emph{IEEE
  Transactions on Affective Computing}, vol.~5, no.~2, pp. 201--213, 2014.

\bibitem{Golan15-TCM}
S.~B.-C. Ofer~Golan, Yana Sinai-Gavrilov, ``{The Cambridge mindreading
  face-voice battery for children (CAM-C): Complex emotion recognition in
  children with and without autism spectrum conditions},'' \emph{Molecular
  Autism}, vol.~22, no.~6, 2015.

\bibitem{Marchi15-AAS}
E.~Marchi, Y.~Zhang, F.~Eyben, F.~Ringeval, and B.~Schuller, ``{Autism and
  Speech, Language, and Emotion -– a Survey},'' in \emph{{Evaluating the Role
  of Speech Technology in Medical Case Management}}, H.~Patil and
  M.~Kulshreshtha, Eds.\hskip 1em plus 0.5em minus 0.4em\relax Berlin: De
  Gruyter, 2015.

\bibitem{Schuller11-RRE}
B.~Schuller, A.~Batliner, S.~Steidl, and D.~Seppi, ``{Recognising Realistic
  Emotions and Affect in Speech: State of the Art and Lessons Learnt from the
  1st Challenge},'' \emph{Speech Communication, Special Issue on Sensing
  Emotion and Affect -– Facing Realism in Speech Process.}, vol.~53, no.
  9/10, pp. 1062--1087, Nov./Dec. 2011.

\bibitem{lang2007international}
P.~Lang and M.~M. Bradley, ``The international affective picture system (iaps)
  in the study of emotion and attention,'' \emph{Handbook of emotion
  elicitation and assessment}, vol.~29, 2007.

\bibitem{kim2008emotion}
J.~Kim and E.~Andr{\'e}, ``Emotion recognition based on physiological changes
  in music listening,'' \emph{IEEE Transactions on Pattern Analysis and Machine
  Intelligence}, vol.~30, no.~12, pp. 2067--2083, 2008.

\bibitem{mahnob_laughter_db}
S.~Petridis, B.~Martinez, and M.~Pantic, ``The mahnob laughter database,''
  \emph{Image and Vision Computing Journal}, vol.~31, no.~2, pp. 186--202,
  February 2013.

\bibitem{Dhall12-CLR}
A.~Dhall, R.~Goecke, S.~Lucey, and T.~Gedeon, ``{Collecting Large, Richly
  Annotated Facial-Expression Databases from Movies},'' \emph{IEEE Transactions
  on Multimedia}, vol.~19, no.~3, pp. 34--41, 2012.

\bibitem{walter2013transsituational}
S.~Walter, J.~Kim, D.~Hrabal, S.~C. Crawcour, H.~Kessler, and H.~C. Traue,
  ``Transsituational individual-specific biopsychological classification of
  emotions,'' \emph{IEEE Transactions on Systems, Man, and Cybernetics:
  Systems}, vol.~43, no.~4, pp. 988--995, 2013.

\bibitem{nicolaou_tpamiDPCCA}
M.~A. Nicolaou, V.~Pavlovic, and M.~Pantic, ``Dynamic probabilistic cca for
  analysis of affective behaviour and fusion of continuous annotations,''
  \emph{IEEE Transactions on Pattern Analysis and Machine Intelligence},
  vol.~36, no.~7, pp. 1299--1311, 2014.

\bibitem{GCTW}
F.~Zhou and F.~{De la Torre}, ``Generalized canonical time warping,''
  \emph{Transactions on Pattern Analysis and Machine Intelligence (PAMI)},
  vol.~38, no.~2, pp. 279--294, 2016.

\bibitem{kipp2001anvil}
M.~Kipp, ``Anvil-a generic annotation tool for multimodal dialogue,'' in
  \emph{Proc. 7th European Conference on Speech Communication and Technology},
  2001.

\bibitem{meudt2012atlas}
S.~Meudt, L.~Bigalke, and F.~Schwenker, ``Atlas--an annotation tool for hci
  data utilizing machine learning methods,'' \emph{Proc. of APD}, vol.~12, pp.
  5347--5352, 2012.

\bibitem{bock2011ikannotate}
R.~B{\"o}ck, I.~Siegert, M.~Haase, J.~Lange, and A.~Wendemuth, ``ikannotate--a
  tool for labelling, transcription, and annotation of emotionally coloured
  speech,'' in \emph{International Conference on Affective Computing and
  Intelligent Interaction}.\hskip 1em plus 0.5em minus 0.4em\relax Springer,
  2011, pp. 25--34.

\bibitem{scherer2010developing}
S.~Scherer, I.~Siegert, L.~Bigalke, and S.~Meudt, ``Developing an expressive
  speech labelling tool incorporating the temporal characteristics of
  emotion.'' in \emph{LREC}, 2010.

\bibitem{cowie2000feeltrace}
R.~Cowie, E.~Douglas-Cowie, S.~Savvidou*, E.~McMahon, M.~Sawey, and
  M.~Schr{\"o}der, ```feeltrace': An instrument for recording perceived emotion
  in real time,'' in \emph{ISCA tutorial and research workshop (ITRW) on speech
  and emotion}, 2000.

\bibitem{Ringeval13-ITR}
F.~Ringeval, A.~Sonderegger, J.~Sauer, and D.~Lalanne, ``Introducing the
  {RECOLA} multimodal corpus of remote collaborative and affective
  interactions,'' in \emph{Proc. International Workshop on Emotion
  Representation, Analysis and Synthesis in Continuous Time and Space
  (EmoSPACE), FG}, Shanghai, China, 2013, 8 pages.

\bibitem{AFEWVA_kossaifi}
J.~Kossaifi, G.~Tzimiropoulos, S.~Todorovic, and M.~Pantic, ``Afew-va database
  for valence and arousal estimation in-the-wild,'' \emph{Image and Vision
  Computing}, vol.~65, pp. 23 -- 36, 2017, multimodal Sentiment Analysis and
  Mining in the Wild Image and Vision Computing.

\bibitem{RCICA}
Y.~Panagakis, M.~A. Nicolaou, S.~Zafeiriou, and M.~Pantic, ``Robust correlated
  and individual component analysis,'' \emph{IEEE Transactions on Pattern
  Analysis and Machine Intelligence, Special Issue in Multimodal Pose
  Estimation and Behaviour Analysis}, 2016.

\bibitem{automatic_survey_2015}
E.~Sariyanidi, H.~Gunes, and A.~Cavallaro, ``Automatic analysis of facial
  affect: A survey of registration, representation, and recognition,''
  \emph{IEEE Transactions on Pattern Analysis and Machine Intelligence},
  vol.~37, no.~6, pp. 1113--1133, 2015.

\bibitem{scherer1997lost}
K.~R. Scherer and G.~Ceschi, ``Lost luggage: a field study of
  emotion--antecedent appraisal,'' \emph{Motivation and emotion}, vol.~21,
  no.~3, pp. 211--235, 1997.

\bibitem{schiel2002smartkom}
F.~Schiel, S.~Steininger, and U.~T{\"u}rk, ``{The SmartKom Multimodal Corpus at
  BAS.}'' in \emph{LREC}, 2002.

\bibitem{Valstar13-A2T}
M.~Valstar, B.~Schuller, K.~Smith, F.~Eyben, B.~Jiang, S.~Bilakhia,
  S.~Schnieder, R.~Cowie, and M.~Pantic, ``{AVEC 2013 –- The Continuous
  Audio/Visual Emotion and Depression Recognition Challenge},'' in \emph{{Proc.
  3rd ACM international workshop on Audio/Visual Emotion Challenge}},
  ACM.\hskip 1em plus 0.5em minus 0.4em\relax Barcelona, Spain: ACM, Oct. 2013,
  pp. 3--10.

\bibitem{sneddon2012belfast}
I.~Sneddon, M.~McRorie, G.~McKeown, and J.~Hanratty, ``The belfast induced
  natural emotion database,'' \emph{IEEE Transactions on Affective Computing},
  vol.~3, no.~1, pp. 32--41, 2012.

\bibitem{aubrey2013cardiff}
A.~J. Aubrey, D.~Marshall, P.~L. Rosin, J.~Vendeventer, D.~W. Cunningham, and
  C.~Wallraven, ``Cardiff conversation database (ccdb): A database of natural
  dyadic conversations,'' in \emph{Proceedings of the IEEE Conference on
  Computer Vision and Pattern Recognition Workshops}, 2013, pp. 277--282.

\bibitem{Valstar14-A2T}
M.~Valstar, B.~Schuller, K.~Smith, T.~Almaev, F.~Eyben, J.~Krajewski, R.~Cowie,
  and M.~Pantic, ``{AVEC 2014 -- The Three Dimensional Affect and Depression
  Challenge},'' in \emph{{Proc. 4th ACM international workshop on Audio/Visual
  Emotion Challenge}}, ACM.\hskip 1em plus 0.5em minus 0.4em\relax Orlando, FL:
  ACM, Nov. 2014.

\bibitem{vandeventer20154d}
J.~Vandeventer, A.~J. Aubrey, P.~L. Rosin, and D.~Marshall, ``{4D Cardiff
  Conversation Database (4D CCDb): A 4D database of natural, dyadic
  conversations},'' in \emph{Proceedings of the 1st Joint Conference on Facial
  Analysis, Animation and Auditory-Visual Speech Processing (FAAVSP 2015)},
  2015.

\bibitem{labov1972sociolinguistic}
W.~Labov, \emph{Sociolinguistic patterns}.\hskip 1em plus 0.5em minus
  0.4em\relax University of Pennsylvania Press, 1972, no.~4.

\bibitem{wittenburg2006elan}
P.~Wittenburg, H.~Brugman, A.~Russel, A.~Klassmann, and H.~Sloetjes, ``Elan: a
  professional framework for multimodality research,'' in \emph{Proceedings of
  LREC}, vol. 2006, 2006, p. 5th.

\bibitem{Schuller13-TI2}
B.~Schuller, S.~Steidl, A.~Batliner, A.~Vinciarelli, K.~Scherer, F.~Ringeval,
  M.~Chetouani, F.~Weninger, F.~Eyben, E.~Marchi \emph{et~al.}, ``The
  {INTERSPEECH} 2013 computational paralinguistics challenge: social signals,
  conflict, emotion, autism,'' in \emph{{Proc.\ of INTERSPEECH}}.\hskip 1em
  plus 0.5em minus 0.4em\relax Lyon, France: ISCA, 2013, pp. 148--152.

\bibitem{Eyben16-TGM}
F.~Eyben, K.~R. Scherer, B.~W. Schuller, J.~Sundberg, E.~Andr{\'e}, C.~Busso,
  L.~Y. Devillers, J.~Epps, P.~Laukka, S.~S. Narayanan, and K.~P. Truong, ``The
  geneva minimalistic acoustic parameter set {(GeMAPS)} for voice research and
  affective computing,'' \emph{IEEE Transactions on Affective Computing},
  vol.~7, no.~2, pp. 190--202, 2016.

\bibitem{shen2015first}
J.~Shen, S.~Zafeiriou, G.~G. Chrysos, J.~Kossaifi, G.~Tzimiropoulos, and
  M.~Pantic, ``The first facial landmark tracking in-the-wild challenge:
  Benchmark and results,'' in \emph{Computer Vision Workshop (ICCVW), 2015 IEEE
  International Conference on}.\hskip 1em plus 0.5em minus 0.4em\relax IEEE,
  2015, pp. 1003--1011.

\bibitem{asthana2014incremental}
A.~Asthana, S.~Zafeiriou, S.~Cheng, and M.~Pantic, ``Incremental face alignment
  in the wild,'' pp. 1859--1866, 2014.

\bibitem{asthana2015pixels}
A.~Asthana, S.~Zafeiriou, G.~Tzimiropoulos, S.~Cheng, and M.~Pantic, ``From
  pixels to response maps: Discriminative image filtering for face alignment in
  the wild,'' \emph{Pattern Analysis and Machine Intelligence, IEEE
  Transactions on}, vol.~37, no.~6, pp. 1312--1320, 2015.

\bibitem{chrysos2015offline}
G.~G. Chrysos, E.~Antonakos, S.~Zafeiriou, and P.~Snape, ``Offline deformable
  face tracking in arbitrary videos,'' in \emph{Proceedings of the IEEE
  International Conference on Computer Vision Workshops}, 2015, pp. 1--9.

\bibitem{kazemi2014one}
V.~Kazemi and J.~Sullivan, ``One millisecond face alignment with an ensemble of
  regression trees,'' in \emph{Proceedings of the IEEE Conference on Computer
  Vision and Pattern Recognition}, 2014, pp. 1867--1874.

\bibitem{wollmer2011robust}
M.~W{\"o}llmer, E.~Marchi, S.~Squartini, and B.~Schuller, ``Robust multi-stream
  keyword and non-linguistic vocalization detection for computationally
  intelligent virtual agents,'' in \emph{International Symposium on Neural
  Networks}.\hskip 1em plus 0.5em minus 0.4em\relax Springer, 2011, pp.
  496--505.

\bibitem{weninger2011munich}
F.~Weninger, J.~Geiger, M.~W{\"o}llmer, B.~Schuller, and G.~Rigoll, ``The
  munich 2011 chime challenge contribution: Nmf-blstm speech enhancement and
  recognition for reverberated multisource environments.''

\bibitem{walecki2015variable}
R.~Walecki, O.~Rudovic, V.~Pavlovic, and M.~Pantic, ``Variable-state latent
  conditional random fields for facial expression recognition and action unit
  detection,'' in \emph{Proceedings of IEEE International Conference on
  Automatic Face and Gesture Recognition (FG'15)}, Ljubljana, Slovenia, May
  2015, pp. 1--8.

\bibitem{mavadati2013disfa}
S.~M. Mavadati, M.~H. Mahoor, K.~Bartlett, P.~Trinh, and J.~F. Cohn, ``Disfa: A
  spontaneous facial action intensity database,'' \emph{IEEE Transactions on
  Affective Computing}, vol.~4, no.~2, pp. 151--160, 2013.

\bibitem{FERA2015}
M.~F. Valstar, T.~Almaev, J.~M. Girard, G.~McKeown, M.~Mehu, L.~Yin, M.~Pantic,
  and J.~F. Cohn, ``Fera 2015-second facial expression recognition and analysis
  challenge,'' in \emph{Automatic Face and Gesture Recognition (FG), 2015 11th
  IEEE International Conference and Workshops on}, vol.~6.\hskip 1em plus 0.5em
  minus 0.4em\relax IEEE, 2015, pp. 1--8.

\bibitem{shen2013framework}
J.~Shen and M.~Pantic, ``Hci\^2 framework: A software framework for multimodal
  human-computer interaction,'' \emph{Transactions on Cybernetics}, vol.~43,
  no.~6, pp. 1593--1606, 2013.

\bibitem{avec_2012}
B.~Schuller, M.~Valster, F.~Eyben, R.~Cowie, and M.~Pantic, ``{AVEC 2012: the
  continuous audio/visual emotion challenge},'' \emph{Proc. 14th Int'l Conf.
  Multimodal Interaction Workshops}, pp. 449--456, 2012.

\bibitem{avec_2014}
M.~Valstar, B.~Schuller, K.~Smith, T.~Almaev, F.~Eyben, J.~Krajewski, R.~Cowie,
  and M.~Pantic, ``{AVEC 2014: 3D Dimensional Affect and Depression Recognition
  Challenge},'' \emph{Proceedings of the 4th ACM International Workshop on
  Audio/Visual Emotion Challenge (AVEC '14)}, pp. 3--10, 2014.

\bibitem{avec_2016}
M.~Valstar, J.~Gratch, B.~Schuller, F.~Ringeval, D.~Lalanne, M.~Torres~Torres,
  S.~Scherer, G.~Stratou, R.~Cowie, and M.~Pantic, ``Avec 2016: Depression,
  mood, and emotion recognition workshop and challenge,'' in \emph{Proceedings
  of the 6th International Workshop on Audio/Visual Emotion Challenge}, ser.
  AVEC '16.\hskip 1em plus 0.5em minus 0.4em\relax New York, NY, USA: ACM,
  2016, pp. 3--10.

\bibitem{baltrusaitis2013dimensional}
T.~Baltrusaitis, N.~Banda, and P.~Robinson, ``Dimensional affect recognition
  using continuous conditional random fields,'' in \emph{Automatic Face and
  Gesture Recognition Workshops)}.\hskip 1em plus 0.5em minus 0.4em\relax IEEE,
  2013, pp. 1--8.

\bibitem{nicolaou2012output}
M.~A. Nicolaou, H.~Gunes, and M.~Pantic, ``Output-associative rvm regression
  for dimensional and continuous emotion prediction,'' \emph{Image and Vision
  Computing}, vol.~30, no.~3, pp. 186--196, 2012.

\bibitem{scikit-learn}
F.~Pedregosa, G.~Varoquaux, A.~Gramfort, V.~Michel, B.~Thirion, O.~Grisel,
  M.~Blondel, P.~Prettenhofer, R.~Weiss, V.~Dubourg, J.~Vanderplas, A.~Passos,
  D.~Cournapeau, M.~Brucher, M.~Perrot, and E.~Duchesnay, ``Scikit-learn:
  Machine learning in {P}ython,'' \emph{Journal of Machine Learning Research},
  vol.~12, pp. 2825--2830, 2011.

\bibitem{evaluation_supervised_learning}
R.~Caruana, N.~Karampatziakis, and A.~Yessenalina, ``An empirical evaluation of
  supervised learning in high dimensions,'' in \emph{Proceedings of the 25th
  International Conference on Machine Learning}.\hskip 1em plus 0.5em minus
  0.4em\relax New York, NY, USA: ACM, 2008, pp. 96--103.

\bibitem{3D_emotion_random_forest}
H.~Chen, J.~Li, F.~Zhang, Y.~Li, and H.~Wang, ``3d model-based continuous
  emotion recognition,'' in \emph{2015 IEEE Conference on Computer Vision and
  Pattern Recognition (CVPR)}, June 2015, pp. 1836--1845.

\bibitem{Hochreiter97-LST}
S.~Hochreiter and J.~Schmidhuber, ``Long short-term memory,'' \emph{Neural
  computation}, vol.~9, no.~8, pp. 1735--1780, Nov 1997.

\bibitem{Graves12-SSL}
A.~Graves, \emph{Supervised sequence labelling with recurrent neural
  networks}.\hskip 1em plus 0.5em minus 0.4em\relax Berlin/Heidelberg, Germany:
  Springer, 2012, vol. 385.

\bibitem{Zhang16-Facing}
Z.~Zhang, F.~Ringeval, J.~Han, J.~Deng, E.~Marchi, and B.~Schuller, ``Facing
  realism in spontaneous emotion recognition from speech: Feature enhancement
  by autoencoder with {LSTM} neural networks,'' in \emph{Proc.\ INTERSPEECH},
  San Francisco, CA, 2016, pp. 3593--3597.

\bibitem{Han16-strength}
J.~Han, Z.~Zhang, N.~Cummins, F.~Ringeval, and B.~Schuller, ``Strength
  modelling for real-worldautomatic continuous affect recognition from
  audiovisual signals,'' \emph{Image and Vision Computing}, 2016.

\bibitem{Weninger15-ICT2}
F.~Weninger, J.~Bergmann, and B.~Schuller, ``{Introducing CURRENNT: the Munich
  Open-Source CUDA RecurREnt Neural Network Toolkit},'' \emph{J. Machine
  Learning Research}, vol.~16, pp. 547--551, 2015.

\bibitem{sift_lowe}
D.~G. Lowe, ``Distinctive image features from scale-invariant keypoints,''
  \emph{International Journal of Computer Vision (IJCV)}, vol.~60, no.~2, pp.
  91--110, 2004.

\bibitem{bidirectional_aam}
J.~Kossaifi, G.~Tzimiropoulos, and M.~Pantic, ``Fast and exact bi-directional
  fitting of active appearance models,'' in \emph{Proceedings of the IEEE
  Int’l Conf. on Image Processing (ICIP’15)}, Quebec City, QC, Canada,
  September 2015, pp. 1135--1139.

\bibitem{newton_aam}
------, ``Fast newton active appearance models,'' in \emph{Proceedings of the
  IEEE Int’l Conf. on Image Processing (ICIP’14)}, Paris, France, October
  2014, pp. 1420--1424.

\bibitem{aam_tip}
------, ``Fast and exact newton and bidirectional fitting of active appearance
  models,'' \emph{IEEE Transactions on Image Processing (TIP), accepted for
  publication}, 2016.

\bibitem{Eyben16-RSA}
F.~Eyben, \emph{Real-time speech and music classification by large audio
  feature space extraction}.\hskip 1em plus 0.5em minus 0.4em\relax Springer,
  2016.

\bibitem{Eyben13-RDI}
F.~Eyben, F.~Weninger, F.~Gro{\ss}, and B.~Schuller, ``Recent developments in
  {openSMILE}, the munich open-source multimedia feature extractor,'' in
  \emph{{Proc. of the 21st ACM International Conference on Multimedia (ACM
  MM)}}.\hskip 1em plus 0.5em minus 0.4em\relax Barcelona, Spain: ACM, 2013,
  pp. 835--838.

\bibitem{Schuller14-TI2}
B.~Schuller, S.~Steidl, A.~Batliner, J.~Epps, F.~Eyben, F.~Ringeval, E.~Marchi,
  and Y.~Zhang, ``The {INTERSPEECH} 2014 computational paralinguistics
  challenge: cognitive \& physical load.'' in \emph{{Proc.\ of
  INTERSPEECH}}.\hskip 1em plus 0.5em minus 0.4em\relax Singapore, Singapore:
  ISCA, 2014, pp. 427--431.

\bibitem{Schuller15-TI2}
B.~Schuller, S.~Steidl, A.~Batliner, S.~Hantke, F.~H{\"o}nig, J.~R.
  Orozco-Arroyave, E.~N{\"o}th, Y.~Zhang, and F.~Weninger, ``The interspeech
  2015 computational paralinguistics challenge: Nativeness, parkinson \& eating
  condition,'' in \emph{{Proc.\ of INTERSPEECH}}.\hskip 1em plus 0.5em minus
  0.4em\relax Dresden, Germany: ISCA, 2015, pp. 478--482.

\bibitem{Ringeval15-A2T}
F.~Ringeval, B.~Schuller, M.~Valstar, S.~Jaiswal, E.~Marchi, D.~Lalanne,
  R.~Cowie, and M.~Pantic, ``{AV+EC} 2015: The first affect recognition
  challenge bridging across audio, video, and physiological data,'' in
  \emph{{Proc.\ of the 5th International Workshop on Audio/Visual Emotion
  Challenge (AVEC)}}.\hskip 1em plus 0.5em minus 0.4em\relax Brisbane,
  Australia: ACM, 2015, pp. 3--8.

\bibitem{russel2003}
J.~Russell, B.~J.A, and J.~Fernandez-Dols, ``Facial and vocal expressions of
  emotions,'' \emph{Annu. Rev. Psychol.}, vol.~54, pp. 329--349, 2003.

\bibitem{Kroschel07}
M.~Grimm and K.~Kroschel, ``Emotion estimation in speech using a 3d emotion
  space concept,'' in \emph{Robust Speech}, M.~Grimm and K.~Kroschel,
  Eds.\hskip 1em plus 0.5em minus 0.4em\relax Rijeka: IntechOpen, 2007, ch.~16.

\end{thebibliography}

\vspace{-20pt}

\begin{IEEEbiography}[{\includegraphics[width=1in,height=1.25in,clip,keepaspectratio]{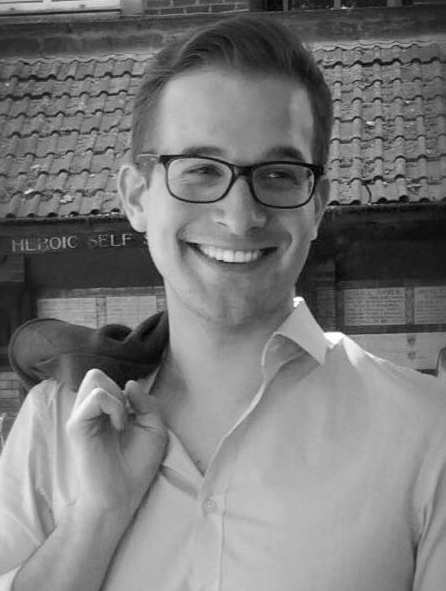}}]{Jean Kossaifi}
is a Research Scientist at Samsung AI Cambridge and a Research Associate at Imperial College London. Bridging the gap between computer vision and machine learning, his research primarily focuses on face analysis and facial affect estimation in naturalistic conditions. Jean is currently working on tensor methods, and how to efficiently combine these with deep learning to develop models which are memory and computationally efficient, and more robust to noise and domain shift. He has co-organized several workshops and tutorials in conjunction with international conferences and is the creator of TensorLy, a high-level API for tensor methods and deep tensorized neural networks in Python that aims at making tensor learning simple and accessible. Jean received his PhD from Imperial College London, where he also previously obtained an MSc in advanced computing. He also holds a French engineering diploma and a BSc in advanced mathematics.
\end{IEEEbiography}
\vspace{-15pt}

\begin{IEEEbiography}
[{\includegraphics[width=1in,height=1.25in,clip,keepaspectratio]{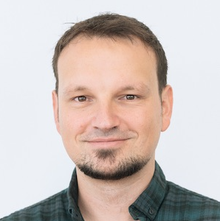}}]
{Robert Walecki} 
received his Ph.D degree at the Department of Computing, Imperial College London, UK. His research interests span the areas of Computer Vision and Machine Learning with application to automatic human behavior analysis. 
He is currently working as a Senior Researcher at the AI team at Babylon Health. His work covers research and applications of a wide range of machine-learning techniques with the goal to put an affordable and accessible health service in the hands of every person on earth.\end{IEEEbiography}
\vspace{-15pt}

\begin{IEEEbiography}[{\includegraphics[width=1in,height=1.25in,clip,keepaspectratio]{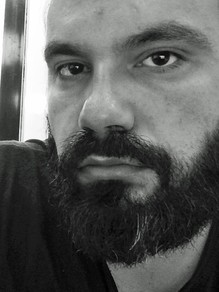}}]{Yannis Panagakis}
is an Associate Professor in machine learning and signal processing at the University of Athens, Department of Informatics and Telecommunications. Prior to this appointment in 2019, he has held research and academic positions at Samsung AI Centre in Cambridge, Imperial College, and Middlesex University in London. His research interests lie in algorithmic and applied aspects of machine learning and its interface with signal processing, high-dimensional and robust statistics, and mathematical optimization.
Dr Panagakis is currently the Managing Editor of the Image and Vision Computing Journal (Elsevier). He co-organized the BMVC 2017 conference and several workshops and special sessions in conjunction with international conferences such as ICCV, EUSIPCO, etc. Yannis received his PhD and MSc degrees from the Department of Informatics, Aristotle University of Thessaloniki and his BSc degree in Informatics and Telecommunication from the University of Athens, Greece.
\end{IEEEbiography}
\vspace{-15pt}

\begin{IEEEbiography}[{\includegraphics[width=1in,height=1.25in,clip,keepaspectratio]{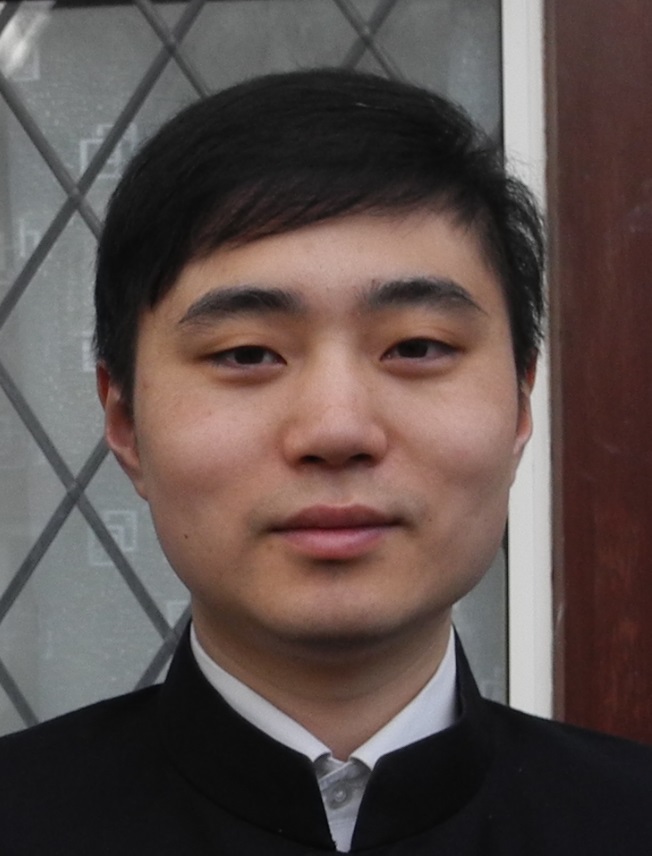}}]{Jie Shen}
 is a researcher at Samsung AI Centre Cambridge and a research fellow at Department of Computing, Imperial College London. He received B.Eng. in electronic engineering from Zhejiang University in 2005, and Ph.D. in computing from Imperial College London in 2014. His current research focuses on affect-sensitive human-computer/human-robot interaction, face parsing, face re-id, and eye gaze analysis.
\end{IEEEbiography}
\vspace{-15pt}

\begin{IEEEbiography}[{\includegraphics[width=1in,height=1.25in,clip,keepaspectratio]{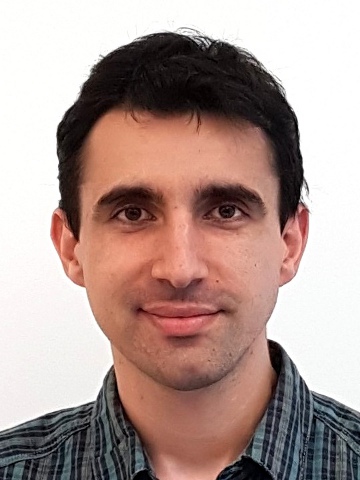}}]{Maximilian Schmitt}
is a PhD student at the ZD.B Chair of Embedded Intelligence for Health Care and Wellbeing at the University of Augsburg, Germany. He received his diploma degree (Dipl.-Ing.) in Electrical Engineering and Information Technology from RWTH Aachen University, Germany, in 2012. Previously, he has worked for the University of Music Detmold and the Computer Science department of the University of Passau, both in Germany. His research focus is on signal processing, machine learning, intelligent audio analysis, and multimodal affect recognition. He has (co-)authored more than 50 publications leading to more than 500 citations (h-index=10). 
He has served as a reviewer for IEEE Transactions on Affective Computing (T-AffC), IEEE Signal Processing Letters (SPL), IEEE Transaction on Cybernetics, IEEE Transactions on Neural Networks and Learning Systems (TNNLS), Elsevier Computer Speech \& Language (CSL), and Elsevier Knowledge-Based Systems (KNOSYS). 
\end{IEEEbiography}
\vspace{-15pt}

\begin{IEEEbiography}[{\includegraphics[width=1in,height=1.25in,clip,keepaspectratio]{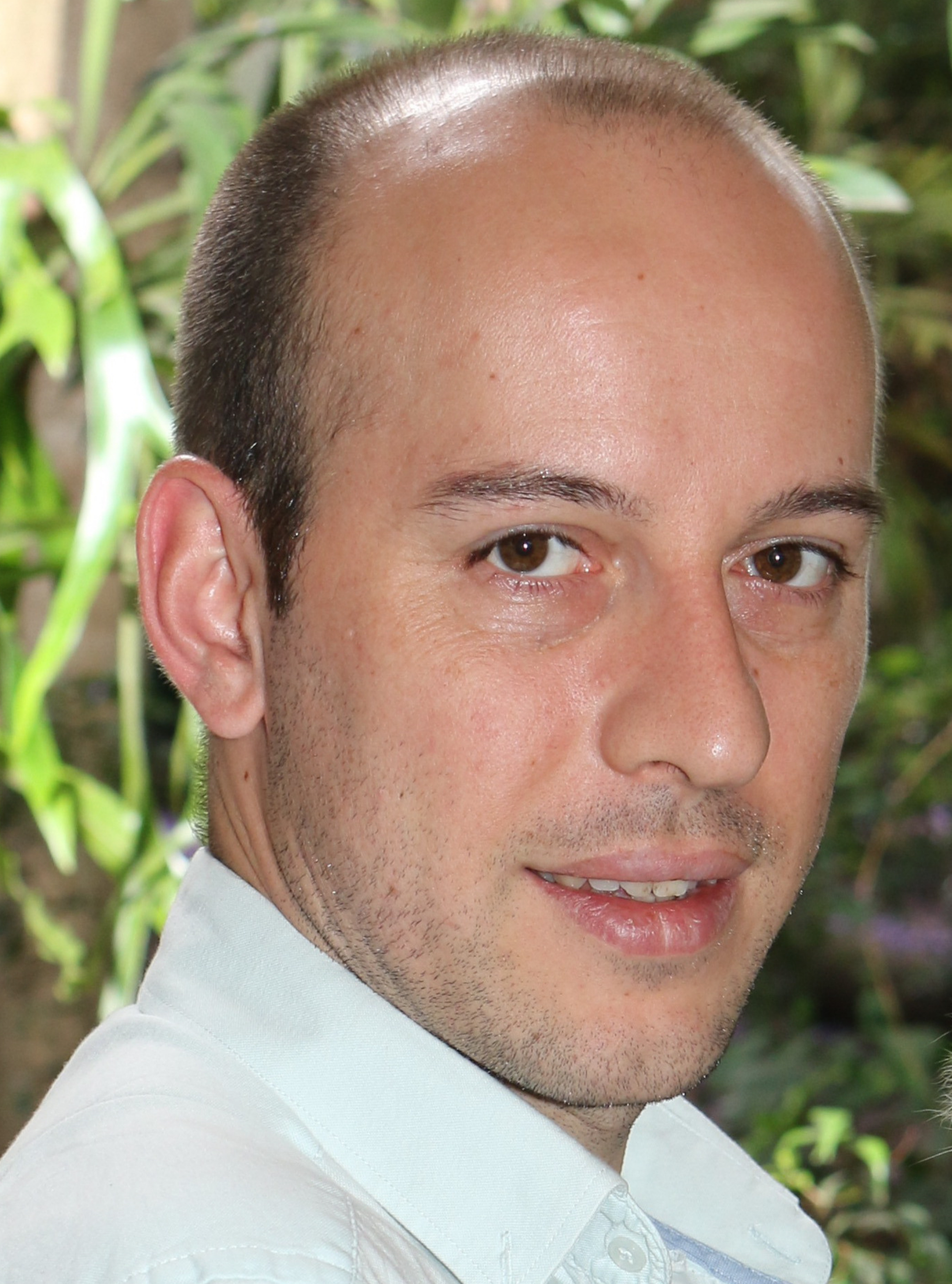}}]{Fabien Ringeval} 
received the M.S. degree in speech and image signal processing from the Universit\'e Pierre et Marie Curie (UPMC), Paris, France, in 2006 and the Ph.D. degree for his researches on the automatic recognition of acted and spontaneous emotions from speech in 2011, from the same university. He is Associate Professor in the team GETALP at the Laboratoire d'Informatique de Grenoble (LIG), CNRS, Université Grenoble Alpes, France, since 2016. His research interests concern digital signal processing and machine learning, with applications on the automatic recognition of paralinguistic information (e.g., emotions, social and atypical behaviours) from multimodal data (e.g., audio, video and physiological signals), at the cross-road of computer sciences and human behaviour understanding. Dr. Ringeval (co-)authored more than 70 publications in peer-reviewed books, journals and conference proceedings in the field leading to more than 2,600 citations (h-index = 21). He co-organised several workshops and international challenges, including the International Audio/Visual Emotion Challenge and Workshop (AVEC) series since 2015, and the INTERSPEECH 2013 ComParE challenge. He served as Area Chair for ACM MM 2019, Senior Program Committee member for ICMI 2019, and ACII 2019, Grand Challenge Chair for ICMI 2018, Publication Chair for ACII 2017, and acts as a reviewer for funding and projects (ANR, NSERC, NWO), as well as several IEEE journals and other leading international journals, conferences and workshops in the field.
\end{IEEEbiography}

\vspace{-10pt}
\begin{IEEEbiography}[{\includegraphics[width=1in,height=1.25in,clip,keepaspectratio]{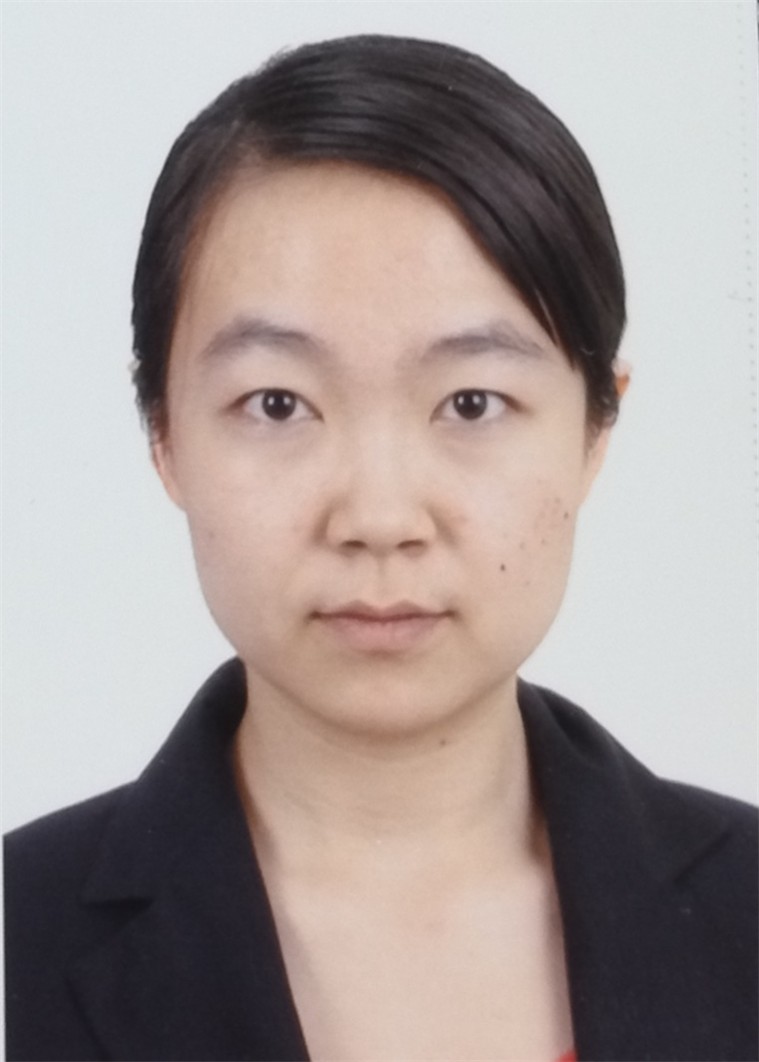}}]{Jing Han}
(S'16) received her bachelor degree (2011) in electronic and information engineering from Harbin Engineering University (HEU), China, and her master degree (2014) from Nanyang Technological University, Singapore. She is now working as a doctoral student with the ZD.B Chair of Embedded Intelligence for Health Care and Wellbeing at the University of Augsburg, Germany, involved in two EU's Horizon 2020 projects SEWA and RADAR CNS. She reviews regularly for IEEE Transactions on Cybernetics, IEEE Transactions on Affective Computing, IEEE Transactions on Multimedia, IEEE Journal of Selected Topics in Signal Processing, and others. Her research interests are related to deep learning for multimodal affective computing and health care. Besides, she co-chaired the 7th Audio/Visual Emotion Challenge (AVEC) and workshop in 2017, and served as a program committee member of the 8th AVEC challenge and workshop in 2018. Moreover, she was awarded student travel grants from IEEE SPS and ISCA to attend ICASSP and INTERSPEECH in 2018.
\end{IEEEbiography}
\vspace{-10pt}

\begin{IEEEbiography}[{\includegraphics[width=1in,height=1.25in,clip,keepaspectratio]{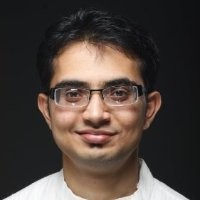}}]
{Vedhas Pandit} (S'11) received his bachelor degree (2008) in electrical engineering from College of Engineering Pune, India, and his master degree (2010) from Arizona State University, USA. He is currently a PhD student at the ZD.B Chair of Embedded Intelligence for Health Care and Wellbeing at the University of Augsburg, Germany. He has regularly served as a reviewer for IEEE Transactions on Neural Networks and Learning Systems (TNNLS), IEEE Journal of Biomedical and Health Informatics (JBHI), IEEE Transactions on Affective Computing (T-AffC),
IEEE Signal Processing Letters (SPL), IEEE Access, Computer Speech and
Language Journal (CSL), among many other journals and conferences. He
has in his name, a novel graphical technique for discrete logic
representation and optimisation, called the `Pandit-Plot'.  His active
research interests and contributions span a wide range of fields; namely
semiconductor device characterisation, music information retrieval,
computer vision, affective computing, model explainability, statistics,
combinatorics, discrete logic systems and cellular automata.
\end{IEEEbiography}
\vspace{-10pt}

\begin{IEEEbiography}[{\includegraphics[width=1in,height=1.25in,clip,keepaspectratio]{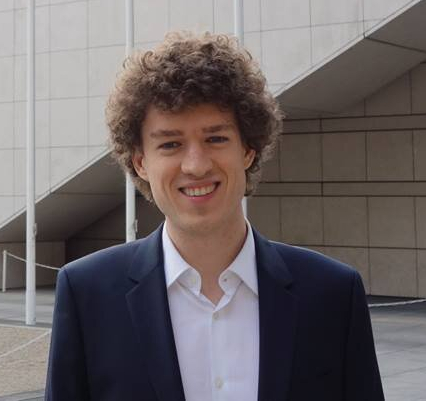}}]{Antoine Toisoul}
is currently a research scientist at Samsung AI with an interest in machine learning applied to computer vision and graphics. He did his PhD in computer graphics at Imperial College London in the EPSRC Centre for Doctoral Training in High Performance Embedded and Distributed Systems (HiPEDS) and the Realistic Graphics and Imaging group under the supervision of Dr Abhijeet Ghosh. In 2014, he received a joint MSc degree between T\'el\'ecom ParisTech (previously known as ENST) and Imperial College London.
\end{IEEEbiography}
\vspace{-10pt}

 \begin{IEEEbiography}[{\includegraphics[width=1in,height=1.25in,clip,keepaspectratio]{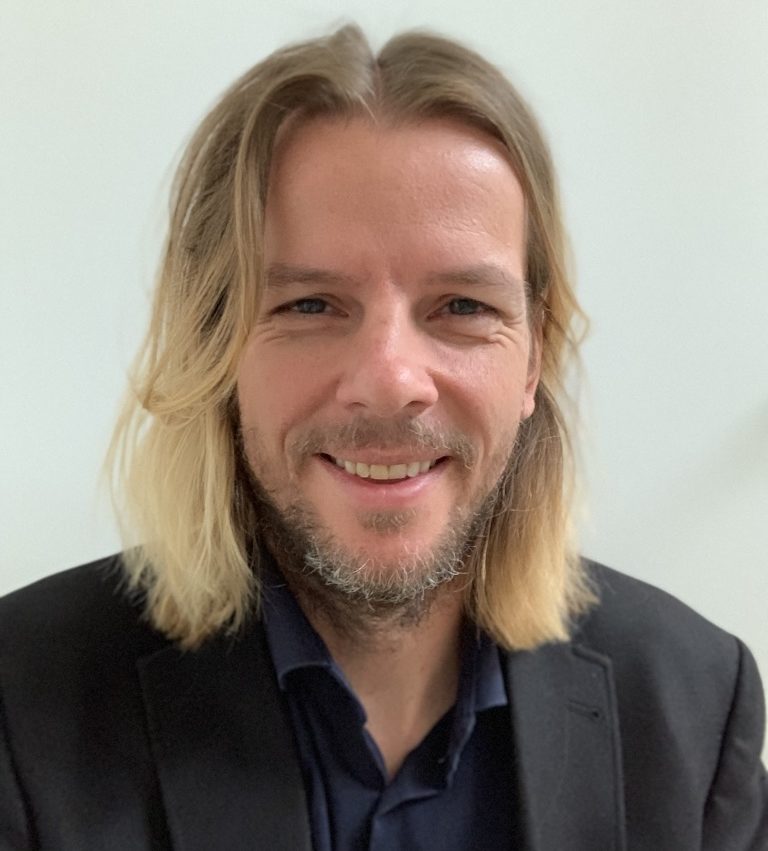}}]{Bj\"orn Schuller}
(M'06, SM'15, Fellow'18) received the Diploma in 1999, the Doctoral degree in 2006, and the Habilitation and Adjunct Teaching Professorship in the subject area of signal processing and machine intelligence in 2012, all in electrical engineering and information technology from Technische Universit\"at M\"unchen, Munich, Germany. He is Professor of Artificial Intelligence in the Department of Computing, Imperial College London, U.K., a Full Professor and head of the ZD.B Chair of Embedded Intelligence for Health Care and Wellbeing at the University of Augsburg, Germany. He (co-)authored five books and more than 800 publications in peer reviewed books, journals, and conference proceedings leading to more than 25\,000 citations (h-index = 74).
\end{IEEEbiography}     
\vspace{-10pt}

\begin{IEEEbiography}[{\includegraphics[width=1in,height=1.25in,clip,keepaspectratio]{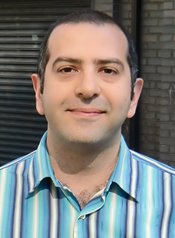}}]{Kam Star} is a digital media entrepreneur, inventor, researcher, investor and award winning games developer. Creating his first computer game in 1986, Kam studied architecture before completing his PhD on games and gamification. He is deeply passionate about innovation in play, behavioural influence and artificial intelligence.
\end{IEEEbiography}

\begin{IEEEbiography}[{\includegraphics[width=1in,height=1.25in,clip,keepaspectratio]{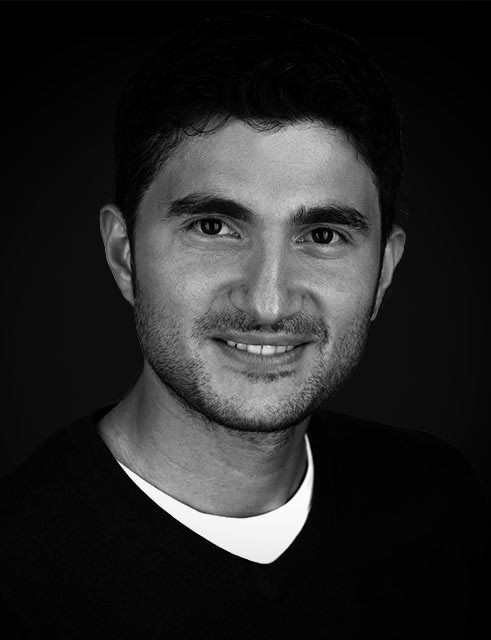}}]{Elnar Hajiyev} 
leads the R\&D and engineering teams at Realeyes, specifically focusing on company advancements in Machine Learning and Big Data areas. He holds BSc (first class) in Information and Media Technology from Brandenburg University of Technology, Germany and PhD in Computer Science from University of Oxford. During his education he received numerous excellency awards, including an Overseas Research Student award, scholarships from Shell and DAAD. Elnar has significant experience working in high-technology companies, including the largest Internet Service Provider of Azerbaijan, Siemens Magnet Technology and his previous technology startup, Semmle, who specialise in on-demand software analytics. He has published more than ten peer-reviewed papers in international journals and conference proceedings, has 5 granted and dozens pending patents internationally.
\end{IEEEbiography}

\begin{IEEEbiography}[{\includegraphics[width=1in,height=1.25in,clip,keepaspectratio]{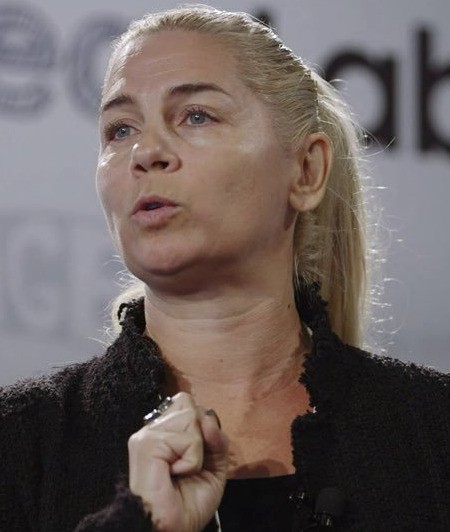}}]{Maja Pantic} 
is a professor in affective and behavioral computing in
the Department of Computing at Imperial College London, UK, and the
research director of Samsung AI Research Centre in Cambridge, UK. She
currently serves as an associate editor for both the IEEE Transactions
on Pattern Analysis and Machine Intelligence and the IEEE Transactions
on Affective Computing. She has received various awards for her work
on automatic analysis of human behavior, including the Roger Needham
Award 2011. She is a fellow of the UK's Royal Academy of Engineering,
the IEEE, and the IAPR.
\end{IEEEbiography}

\end{document}